\documentclass{aa}  

\usepackage{graphicx}
\usepackage{txfonts}
\usepackage[colorlinks=true, linkcolor=blue, citecolor=blue, filecolor=blue, urlcolor=blue]{hyperref}
\usepackage{booktabs}
\usepackage{xspace}
\usepackage{amsmath}

\newcommand{\logteff}{\ensuremath{\log\left(T_\mathrm{eff}/\mathrm{K}\right)}\xspace}
\newcommand{\logage}{\ensuremath{\log\left(\mathrm{age}/\mathrm{yr}\right)}\xspace}
\newcommand{\logmini}{\ensuremath{\log\left(M_\mathrm{ini}/\mathrm{M}_\odot\right)}\xspace}
\newcommand{\logmcurr}{\ensuremath{\log\left(M_\mathrm{curr}/\mathrm{M}_\odot\right)}\xspace}
\newcommand{\logl}{\ensuremath{\log\left(L/\mathrm{L}_\odot\right)}\xspace}
\newcommand{\logg}{\ensuremath{\log\left(g/\mathrm{cms}^{-2}\right)}\xspace}
\newcommand{\teff}{\ensuremath{T_\mathrm{eff}}\xspace}
\newcommand{\mini}{\ensuremath{M_\mathrm{ini}}\xspace}
\newcommand{\mcurr}{\ensuremath{M_\mathrm{curr}}\xspace}
\newcommand{\msun}{\ensuremath{\mathrm{M}_\odot}\xspace}
\newcommand{\Grp}{\ensuremath{\mathrm{G}_\mathrm{RP}}\xspace}
\newcommand{\Gbp}{\ensuremath{\mathrm{G}_\mathrm{BP}}\xspace}
\newcommand{\SNRPi}{\ensuremath{\pi/\Delta\pi}\xspace}

\begin{document}

\title{Characterising stellar photometry with invertible neural networks: Joint inference of stellar parameters and extinction}
\titlerunning{cINN inference of stellar parameters and extinction}

\author{Victor F. Ksoll
      \inst{1}
      \and
      Isabelle Joncour\inst{2}
      \and
      Prasanta Gorai\inst{1}
      \and
      Ralf S.\ Klessen\inst{1}
      %\and 
      %others?
      } 

\institute{Zentrum für Astronomie der Universität Heidelberg, Institut für Theoretische Astrophysik, Albert-Ueberle-Str. 2, D-69120 Heidelberg, Germany\\
          \email{v.ksoll@uni-heidelberg.de}
     \and
         Institut de Planétologie et d'Astrophysique de Grenoble, 414 Rue de la Piscine, 38400 Saint-Martin-d'Hères, France\\
         }

\date{Received \today; accepted }

% \abstract{}{}{}{}{}
% 5 {} token are mandatory
 
  \abstract
   {}
   {We introduce an extension to a simulation-based deep-learning approach for the characterisation of stellar photometry and investigate the feasibility of jointly estimating stellar properties and extinction.}
   {We used the PARSEC stellar evolutionary models to train a conditional invertible neural network (cINN) for the purpose of jointly estimating full posterior distributions of stellar parameters and extinction.
   We tested our cINN on synthetic data and real Gaia+2MASS photometry of the two nearby star-forming regions Taurus and Upper Scorpius.
   We also compared our predicted extinction maps to two independent catalogues based on Gaia XP low-resolution spectra.}
   {On the synthetic test data, our trained cINN exhibited a good predictive performance, achieving root mean squared errors (RMSEs) for the maximum a posteriori point estimates of about $0.16$\,mag for $A_V$, $0.07$\,dex for \logteff, $0.16$\,dex for \logg, $0.2$\,dex for \logl and $0.1$\,dex for \logmini. 
   Only stellar age has proven harder to constrain with an RMSE of $0.56$\,dex in \logage.  
   On the real Gaia+2MASS observations of Taurus and Upper Scorpius, we retrieved reasonable extinction maps that exhibit structures that are visually consistent with local optical dark clouds.
   Compared to the independent XP-spectra-based extinction studies, we found a good qualitative agreement and recovered the same prominent spatial structures.
   Quantitatively, however, we identified a systematic overestimation of extinction in our approach of $0.5-0.7$ mag on average compared to the spectroscopic studies.
   We investigated the potential causes of these offsets and identified photometric and distance uncertainties, which are currently unaccounted for in our approach, as likely candidates.
   }
   {We demonstrated the feasibility of jointly estimating stellar parameters and extinction with our cINN approach, making it a promising avenue to further pursue for robust and efficient analysis of large photometric surveys.
   }


   \maketitle
\nolinenumbers
%
%________________________________________________________________

\section{Introduction}
\label{sec:intro}
Over the last decades, large, high-precision photometric surveys have delivered an unprecedented stellar census of our Milky Way and the nearby Universe. 
Surveys such as the Gaia mission \citep{Gaia2016}, the Sloan Digital Sky Survey \citep[SDSS;][]{York2000, Aihara2011, Kollmeier2026SDSS}, or the Pan-STARRS1 Survey \citep{Chambers2016_PanSTARRS1, Flewelling2020} have fundamentally revolutionised our understanding of the Galactic structure by providing photometry for billions of objects, while the high-spatial resolution of the Hubble Space Telescope \citep[HST;][]{Lallo2012} and the James Webb Space Telescope \citep[JWST;][]{Gardner2023} continue to push our high-precision imaging capabilities for stellar populations from the UV to the infrared regime.
Ongoing (and future) surveys, such as the Legacy Survey of Space and Time \citep[LSST;][]{Ivezic2019_LSST} at the Vera C. Rubin Observatory in Chile, which will obtain photometry for 40 billion objects, or the Euclid mission \citep{EuclidCollaboration2025} are bound to even further enlarge this enormous wealth of photometric survey data by orders of magnitude.

Comparing photometric observations to theoretical stellar evolution models is one of the primary analysis pathways to characterise the observed stellar populations \citep{Salaris2005, Kippenhahn2012, VallsGabaud2014, Carroll_Ostlie_2017}.
With the somewhat limited information that broadband photometry can provide, this is, however, not necessarily a trivial task, as the inference of the fundamental stellar properties can often be ambiguous.
Photometric uncertainties, unresolved binarity, intrinsic overlaps in stellar evolution (in e.g.~the Hertzsprung-Russel Diagram), and effects like extinction can induce degeneracies in the inference process \citep{Pont2004, Jorgensen2005, BailerJones2011,  Ksoll2020}.
Extinction, the attenuation and reddening of stellar light from intervening gas and dust between the source and the observer, can be a particularly significant source of uncertainty when characterising stars \citep[see e.g.~][]{BailerJones2011, Howes2019, OCallaghan2024}.
On the other hand, constraining extinction can provide crucial insights into the internal structure of star-forming regions and the distribution of gas and dust in the interstellar medium in general \citep[see e.g.~][]{Lallement2019, Leike2020, Lallement2022, Vergely2022, Edenhofer2024}.

A wide range of methods has been developed to infer extinction, alone or jointly with the stellar parameters, from photometric and low-resolution spectroscopic surveys.
We provide a dedicated review in Section~\ref{sec:literature_overview}. 
Model-based Bayesian approaches provide full posterior distributions but at a high computational cost per source, whereas data-driven and machine-learning approaches are fast but mostly provide point estimates and of-
ten rely on spectra. Simulation-based inference with normalising flows combines the advantages of both: once trained on a
physical forward model, such networks return full posterior distributions for any new observation at negligible cost.

In this study, we present an extension to the simulation-based approach of \cite{Ksoll2020}, which employed a normalising flow architecture to predict full posterior distributions of stellar parameters from HST photometry, when extinction of each star is known (see also Sect.~\ref{sec:literature_normalising_flows}).
Specifically, we lift the requirement of needing individual stellar extinction measurements as an input and investigate the feasibility of simultaneously estimating the stellar parameters and extinction based only on photometric observations.
For this purpose, we revise our training data set generation by implementing a more flexible interpolation scheme for the PARSEC stellar evolutionary models \citep{Bressan2012, Bressan2013, Tang2014,Chen2014,Chen2015} and an on-the-fly photometry synthetisation to increase the fidelity of the synthetic observations during training of our model.
Using a combination of Gaia and 2MASS photometry for the nearby star-forming regions Taurus and Upper Scorpius, we then compare with two independently derived extinction studies that are based on Gaia XP low-resolution spectra \citep[i.e.][]{AndraeEtAl2023_GSP_Phot, Zhang2025} to ascertain whether our photometry-only approach can reasonably approximate their results, using much less information as an input and jointly estimating the stellar properties in a single step.

The remainder of this study is structured as follows. 
Section~\ref{sec:literature_overview} reviews existing approaches to infer extinction and stellar parameters.
In Section~\ref{sec:data_training}, we describe our forward model, that is our interpolation scheme for the PARSEC evolutionary tracks, the synthetic photometry and the sampling of the training data. 
Section~\ref{sec:data_obs} presents the Gaia and 2MASS data of Taurus and Upper Scorpius and the two comparison catalogues.
Section~\ref{sec:method} summarises the cINN architecture, the training procedure, and the derivation of point estimates. 
In Section~\ref{sec:results_synthetic}, we validate the method on synthetic data and in Sect.~\ref{sec:results_taurus_usco}, we present the application to Taurus and Upper Scorpius, including the comparison with the two independent catalogues.
Section~\ref{sec:discussion} discusses current limitations of our approach. 
Lastly, Section~\ref{sec:summary} summarises our conclusions.  

\section{Inference of extinction and stellar parameters - existing approaches}
\label{sec:literature_overview}

In this section, we review the main families of methods used to infer interstellar extinction from large photometric and spectroscopic surveys, alone or together with the stellar parameters.

\subsection{Colour-excess and isochrone-projection methods}
\label{sec:literature_classic}

One classical technique for inferring interstellar extinction from photometric observations is the analysis of the near infrared colour excess (NICE, \citealt{Lada1994}; NICER, \citealt{Lombardi2001}; NICEST, \citealt{Lombardi2009}), where the photometry of reddened stars is compared to unextincted control fields.
More recently, this technique has been enhanced by an unsupervised machine learning approach in the form of Gaussian mixture modelling to enable a probabilistic treatment with applications to VIRCAM and 2MASS photometry \citep[PNICER;][]{Meingast2017}.
With subsequent refinement, this method has produced 2D extinction maps for the VISTA Variables in the Vía Láctea \citep[VVV;][]{Minniti2010_VVV} survey using VIRCAM photometry \citep[XPNICER;][]{Zhang2022} and for the Galactic plane based on photometry from The UKIRT Infrared Deep Sky Survey \citep[UKIDSS;][]{Lawrence2007_UKIDSS} in \cite{Zhang2025b}.
In another recent work, \cite{Kordopatis2023} used classical isochrone fitting/projection based on evolutionary models to derive extinction and stellar properties given photometry from Gaia and The Two Micron All Sky Survey \citep[2MASS;][]{Skrutskie2006A_2MASS}, as well as spectroscopic estimates of stellar atmospheric parameters. 

\subsection{Bayesian inference}
\label{sec:literature_bayesian}

The estimation of extinction (and stellar properties) from photometry has also seen extensive work that employs Bayesian statistics to model full posterior distributions and account for uncertainties and degeneracy in this inverse problem.
These approaches typically use Markov Chain Monte Carlo \citep[MCMC; e.g.\,][]{Gelfand1990} techniques to sample from the target posterior distributions and include applications to SDSS and UKIDSS photometry in \cite{Hanson2014}, as well as a comprehensive analysis \citep[Bayestar;][]{Green2014, Green2015, Green2018} of observations from Pan-STARRS1 and 2MASS that was later complemented with Gaia parallaxes to produce 3D extinction maps \citep{Green_2019}.

Another Bayesian approach called StarHorse \citep{Santiago2016_StarHorse, Queiroz2018, Queiroz2020} used large grids of stellar evolutionary models, on which the posterior probability densities are computed explicitly instead of the MCMC approach, to jointly estimate extinction and stellar properties.
Although the initial StarHorse approach required spectroscopically derived stellar atmospheric parameters in the fitting process, it has also been adapted to derive extinction maps based on purely photometric data from Gaia DR2 \citep{Gaia2018}, Pan-STARRS1, 2MASS, and AllWISE \citep{Cutri2013_AllWISE} in \cite{Anders2019}.
Later, \cite{Anders2022} applied StarHorse to Gaia EDR3 parallaxes and photometry in combination with 2MASS, Pan-STARRS1, SkyMapper \citep{Onken2019}, and AllWISE data to provide updated distances, extinction and stellar parameters for Gaia sources brighter than $m_G=18.5$ mag. 
More recently, a similar grid-based Bayesian approach called BRUTUS \citep{Speagle2025_BRUTUS} has been employed to derive a 3D map extinction for the southern Galactic plane using photometry from the Dark Energy Camera Plane Survey 2 \citep[DECaPS2;][]{Saydjari2023}, VVV survey, 2MASS and “Unofficial” Wide-field Infrared Survey Explorer \citep[UnWISE;][]{Schlafly2019} in combination with Gaia DR3 parallaxes \citep{Zucker2025}.
Other widely used Bayesian tools based on stellar evolutionary models include PARAM \citep{Rodrigues2014}, UniDAM \citep{Mints2017}, BASTA \citep{BorsenKoch2022} and the isochrone-fitting approach of \cite{Sanders2018}.

\subsection{Machine learning and data-driven approaches}
\label{sec:literature_ml}

With the increasing size of modern observational datasets, machine learning techniques have also become more prevalent in estimating extinction and stellar parameters.
For Gaia DR2, for example, \cite{Andrae2018} employed the tree ensemble approach ExtraTrees \citep[extremely randomised trees;][]{GeurtsEtAl2006_ExtraTrees} to infer stellar properties and extinction.
Subsequently, \cite{AndraeEtAl2023_GSP_Phot} used ExtraTrees to initialise and speed-up a Bayesian MCMC approach to derive refined predictions for extinction and stellar parameters based on Gaia spectra and photometry in the GSP-Phot pipeline for Gaia DR3.
\cite{Chen2019MNRAS.483.4277C} trained random forests on spectroscopic reddening estimates to derive reddening values from Gaia, 2MASS and WISE photometry for three-dimensional dust maps of the Galactic plane.
\cite{Green2021} developed a neural network to predict the absolute magnitudes of stars in Gaia, Pan-STARRS1, 2MASS and WISE as a function of their atmospheric parameters, jointly with a passband-dependent reddening vector.
\cite{ZhangEtAl2023_Gaia_BayesModel_Data_APs} integrated an empirical neural network forward model for Gaia spectra, as well as Gaia, 2MASS and WISE photometry, in a Bayesian posterior optimisation framework to estimate stellar atmospheric parameters, distance and extinction and, thus, provide a 3D extinction map.
Later, \cite{Edenhofer2024} used these results to derive a detailed Galactic map of the distribution of dust within $1.25$\,kpc of the Sun.
More recently, \cite{Zhang2025} updated the \cite{ZhangEtAl2023_Gaia_BayesModel_Data_APs} extinction map by revisiting their extinction prediction while leaving the slope $R_V$ of the adopted reddening law as a free parameter.
In other recent data-driven studies, \cite{Khalatyan2024} transferred spectroscopic labels, including $A_V$ and stellar mass, to $217$ million Gaia DR3 sources with gradient-boosted trees using XP spectra, Gaia, 2MASS and All-WISE photometry and parallaxes, and \cite{Leung2024} trained a transformer-based model to predict stellar parameters and reddening from any subset of XP spectra, colours and spectroscopic labels.

\subsection{Simulation-based inference with normalising flows}
\label{sec:literature_normalising_flows}

Grid-based Bayesian approaches can be limited by the coverage of their model grids, while the MCMC-based methods often suffer from high computational costs during inference.
As a computationally efficient alternative to the Bayesian MCMC paradigm, generative deep learning approaches have been explored to directly model and subsequently sample from the target posterior distributions.
These generative approaches belong to the broader class of simulation-based inference methods that learn a model of the posterior (or of the likelihood) from simulations of the forward process \citep{Cranmer2020}. 
Once trained, such estimators provide posterior samples for new observations at negligible cost (i.e.~they are "amortised estimators"), which makes them particularly attractive for large surveys.

Normalising Flows \citep[NFs;][]{Tabak2010, Tabak2013, Dinh2015, Rezende2015, Kobyzev2021, Papamakarios2019}, which model complex distributions by learning bijective mappings to simpler latent distributions, have emerged as a promising avenue in this \citep[see e.g.][for a review]{Kobyzev2021}.
Among the NFs, the conditional invertible neural network \citep[cINN;][]{Ardizzone2019a, Ardizzone2019b} is an architecture that is particularly well-suited for posterior estimation in inverse problems. 
cINNs have already seen a wide array of successful applications in astronomy. 
These include the inference of the physical properties of H{\sc ii} regions from nebular emission-line diagnostics \citep{Kang2022, Kang2023a}, a study of cosmic ray origins \citep{Bister2022}, the characterisation of exoplanet interior structure \citep{Haldemann2023}, retracing the merger histories of galaxies \citep{Eisert2023, Eisert2026}, reconstruction of 3D dust structures in star-forming cores \citep{Ksoll2024}, or the analysis of properties of extragalactic stellar clusters from unresolved photometry \citep{Walter2026}. 
cINNs have also been employed for stellar spectral analysis with simulation-based implementations for MUSE \citep{Kang2023b, Kang2025, Gkimisi2026} or 4MOST spectra \citep{Ksoll2026}, as well as an observation-trained approach that utilised high-resolution GIRAFFE spectra from the Gaia-ESO survey \citep{Candebat2024}.
NFs have recently been applied to the inference of extinction from broad-band photometry as well: \cite{OCallaghan2026} used a conditional NF, trained on observed stars at mid-to-high Galactic latitudes, to infer the reddening and distance of individual stars, and \cite{Kapusta2025} combined a transformer encoder with an NF to infer \teff, log(g), [M/H], $A_V$ and $R_V$ from an arbitrary set of photometric bands.
Lastly, and most relevant to this paper is our previous proof-of-concept study, \cite{Ksoll2020}, where we successfully demonstrated the application of a simulation-based cINN for the purpose of predicting full posterior distributions for stellar properties based on HST photometry. 

\subsection{Positioning of this work}
\label{sec:literature_positioning}

Comparing the existing approaches that provide extinction and stellar parameters for large samples of Gaia sources and that are most similar to ours, two trade-offs emerge.
First, the methods that return full posteriors incur a computational cost that grows with the number of sources, either by sampling (GSP-Phot) or by evaluating a model grid (StarHorse), whereas fast methods summarise the posterior by a point estimate with Gaussian uncertainties \citep{Zhang2025} or by a regression \citep{Khalatyan2024}. 
Second, most of the recent large catalogues rely on the Gaia XP spectra, limiting them to sources with sufficiently bright and well-measured spectra, and only StarHorse provides ages and masses along-side the extinction. 
Recent NF approaches provide amortised posteriors from photometry, but either for the atmospheric parameters and extinction only \citep{Kapusta2025} or for the reddening and distance only \citep{OCallaghan2026}.
Our approach intends to fill this gap: it returns amortised posterior distributions, based on a physical forward model, for the age and initial mass of each star jointly with its luminosity, effective temperature, surface gravity and band-integrated extinction, from broad-band photometry alone.

\section{Forward model and training set}
\label{sec:data_training}

\subsection{Stellar evolutionary models}
\label{sec:data_training_evol}
Building upon \cite{Ksoll2020}, our training set was again based on the PARSEC stellar evolutionary models \citep{Bressan2012, Bressan2013, Chen2014, Chen2015, Chen2019_YBC, Tang2014, Marigo2017, Pastorelli2019, Pastorelli2020,Nguyen2025}, but we revised our training data compilation procedure here.
Previously, we used the PARSEC web-interpolation tool\footnote{\url{http://stev.oapd.inaf.it/cmd}} to obtain isochrone tables, which we then augmented via interpolation to modify the prior distributions of stellar age and mass in the training data. 
Although this procedure worked well enough in \cite{Ksoll2020}, it was inflexible in the choice of the priors (of stellar age and mass) and limited by the framework of the PARSEC web-interpolation tool. 
In particular, in the current web-tool, age can only be sampled (conveniently) on a regular, discretised grid.
While this is perfectly fine for typical use cases of isochrones (such as isochrone fitting), in neural network training it may cause artefacts (e.g.~a network may learn that age is a discrete rather than continuous parameter).
In addition, the PARSEC web-interpolation tool currently only offers the \citet{OD1994} revision of the \citet{CCM1989} galactic extinction law when synthesising photometry.

To improve the flexibility of our approach, we implemented our own PARSEC model interpolator and photometry synthesiser.
For the model interpolation step, we employed version 1.2S of the PARSEC evolutionary tracks\footnote{\url{https://people.sissa.it/~sbressan/parsec.html}} \citep{Bressan2012, Bressan2013, Tang2014,Chen2014,Chen2015}. 
This grid covers stellar masses from $0.09$ to $350\,\mathrm{M}_\odot$ and a metallicity range of $Z=0.0001$ to $Z=0.06$, but does not include the COLIBRI extensions \citep{Marigo2013,Marigo2017,Pastorelli2019, Pastorelli2020}, which model the thermal pulsing asymptotic giant branch (TP-AGB) evolution of the stars. 
As such we excluded the TP-AGB and subsequent phases (such as white dwarves dwarfs) in our analysis. 
To achieve the most flexibility with regards to the sampling of our training data, the primary inputs of our interpolator are metallicity $Z$, stellar age $\log(\mathrm{age})$ and initial stellar mass $M_\mathrm{ini}$. 
Given these inputs, the remaining (secondary) properties of a PARSEC stellar model, such as luminosity, effective temperature, surface gravity, surface abundances or mass loss rate are then interpolated. 
We adopted version 1.2S rather than the more recent PARSEC v2.0 \citep{Nguyen2022} because it is the version underlying GSP-Phot (Appendix~\ref{app:gpspphot}), which simplifies the comparison in Sect.\ref{sec:results_obs_comparison}; an update to the newer models, is planned for our subsequent development.

\subsection{Interpolation of evolutionary tracks}
\label{sec:data_training_interpolation}
Given a query initial mass \mini, our interpolation scheme begins with interpolating a corresponding evolutionary track based on the two neighbouring evolutionary tracks in the PARSEC model grid (i.e.~the next more massive and less massive track, respectively).
For this purpose, we first identified equivalent evolutionary phases (EEPs) along each of the evolutionary tracks in the grid, following a prescription from \cite{Dotter2016} that is also used for the MIST models \citep{Choi2016}.
EEPs represent physically equivalent phases on evolutionary tracks of different (initial) mass and facilitate a physically meaningful interpolation between the tracks. 
\cite{Dotter2016} distinguishes primary and secondary EEPs, where the former denote actual physical phases in the evolution, whereas the latter are equally spaced secondary points (along the track) between the primary EEPs that ensure a smooth interpolation. % between tracks. 
The grid of PARSEC evolutionary tracks does not come in EEP format out of the box, but provides a list of starting points for certain (physically motivated) evolutionary phases for every track. 
We adopted these points as the primary EEPs for the PARSEC evolutionary models and used the approach of \cite{Dotter2016} to populate the secondary EEPs in between.
This procedure defines a distance metric along a given track, where the distance of model $i+1$ is given as a function of the distance of the preceding model $i$ following
\begin{equation}
    D_{i+1} = D_i + \sqrt{\sum_j^N w_j\left(x_{j, i+1} - x_{j, i}\right)^2}, 
\end{equation}
where $D_0 = 0$ and $x_j$ denote individual properties of the model star, whose contribution to the distance metric is weighted by $w_j$. 
The typical properties considered for this metric are effective temperature and luminosity, so that $D$ reflects a distance in the Hertzsprung Russel Diagram (HRD). 
Following \cite{Dotter2016}, we also included the stellar age in the distance computation, such that $x = \left\{\logteff, \logl, \logage\right\}$. 
For the weights, we achieved the best results for $w = \{10, 0.05, 0.001\}$.
After computing these distances for the original sampling of the evolutionary tracks, we regularised all tracks such that they have an equal amount of secondary EEPs between the primary EEPs. 

Following this regularisation, a track for a new query $M_\mathrm{ini}$ is computed via linear interpolation between the corresponding EEPs as a function of the initial mass. 
Subsequently, our interpolator determines the query stellar model by linearly interpolating in $\log(\mathrm{age})$ along the query (interpolated) evolutionary track.
To interpolate in metallicity, we first compute a stellar model (at the query initial mass and age) for the two neighbouring metallicities in the PARSEC grid (i.e.~next higher and next lower metallicity), and then linearly interpolate between these two models to determine the output at the query metallicity.

\subsection{Synthetic photometry}
\label{sec:data_training_photometry}

The PARSEC stellar evolutionary tracks provide only the intrinsic physical properties of the modelled stars.
To train a cINN for the purpose of characterising real photometry, we have to link these properties to synthetic photometric observations. 
This requires the association of each model star to a corresponding synthetic stellar atmosphere model and integrating the respective fluxes convolved with the filter transmission curves for a given set of photometric filters to retrieve the synthetic photometric magnitudes. 
To facilitate a more comprehensive sampling of the extinction than what a pre-computed model grid can offer, we decided to implement this photometry synthetisation step as an on-the-fly data augmentation procedure during training of the cINN.
That means for every batch in every epoch, extinction is randomly sampled from a prescribed prior distribution and the corresponding attenuated broad-band photometry for each training stellar model is computed on-the-fly.
Although this approach slows down training compared to using a pre-computed grid, our implementation (in Python) is fast enough to not do so to a detrimental degree (taking e.g.~about $2$\,ms per star for photometry in six filters on a Intel Xeon Gold 6254 CPU at $3.1$\,GHz).
The following summarises this photometry synthetisation. 

\subsubsection{Computing synthetic magnitudes}
\label{sec:data_training_photometry_magnitudes}
Stellar atmosphere models typically specify the flux $F_\lambda$ of the star at the stellar radius $R$ \citep{Chen2019_YBC}.
From this, the absolute flux $f_{\lambda, 10\mathrm{pc}}$ at a distance of $10$ pc can be derived according to   
\begin{equation}
\label{eq:AbsoluteFlux}
    f_{\lambda, 10\mathrm{pc}} = \left(\frac{R}{10\mathrm{pc}}\right)^2 F_\lambda 10^{-0.4 A_\lambda},
\end{equation}
where $A_\lambda$ denotes the amount of extinction between the observer and the star. 
Following this, the absolute magnitude $M_i$ in a given filter $i$ (for a photon counting detector) is defined as 
\begin{equation}
\label{eq:AbsoluteMagnitude}
\begin{split}
    M_i &= -2.5 \log\left(\frac{\int_{\lambda_1}^{\lambda_2} \lambda f_{\lambda, 10\mathrm{pc}} T_{\lambda, i}\mathrm{d}\lambda}{\int_{\lambda_1}^{\lambda_2} \lambda f_\lambda^0 T_{\lambda, i}\mathrm{d}\lambda} \right) + m_{i, 0} \\
        &= -2.5 \log\left[\left(\frac{R}{10\mathrm{pc}}\right)^2 \frac{\int_{\lambda_1}^{\lambda_2} \lambda F_\lambda 10^{-0.4 A_\lambda} T_{\lambda, i}\mathrm{d}\lambda}{\int_{\lambda_1}^{\lambda_2} \lambda f_\lambda^0 T_{\lambda, i}\mathrm{d}\lambda}\right] + m_{i, 0},
\end{split}
\end{equation}
where $f_\lambda^0$ and $m_{i, 0}$ refer to the reference flux and reference magnitude of the chosen photometric system, respectively, and $T_{\lambda, i}$ describes the filter response curve \citep{Chen2019_YBC}.

\subsubsection{Stellar atmosphere libraries}
\label{sec:data_training_photometry_atmospheres}

To match the stellar evolutionary models with synthetic stellar atmospheres, we adapted the procedure outlined by \cite{Chen2019_YBC} in order to be consistent with the synthetic photometry provided by the PARSEC isochrone web-interpolator. 
\cite{Chen2019_YBC} compiled a catalogue %\footnote{available at \url{https://gitlab.com/cycyustc/spec_ybc}} 
that includes stellar atmospheres from ATLAS 9 \citep{Castelli_Kurucz2003, Kurucz2014}, PHOENIX BT-Settl \citep{AllardHauschildt1995, Allard1997, Allard2012}, WM-basic \citep[O and B stars;][]{Pauldrach1986, Chen2015}, PoWR\footnote{\url{http://www.astro.physik.uni-potsdam.de/~wrh/PoWR/powrgrid1.php}} \citep[Wolf Rayet stars;][]{Graefener2002, Hamann2004, Sander2012, Todt2015}, COMARCS \citep[C-, S-, K-, and M-type TP-AGB stars and other red giants;][]{Aringer2009, Aringer2016, Aringer2019} and Koester \citep[white dwarves;][]{Koester2010, Tremblay2009}. 
From this collection, we adopted all, but the PoWR, COMARCS C-star and Koester WD models. 
We excluded the latter two, because our chosen set of the PARSEC evolutionary tracks does not include the relevant evolutionary phases (i.e.~TP-AGB and white dwarves). 
For the PoWR models, we could not reproduce the output of the PARSEC web-interpolator for the relevant stars based on the implementation details given in \cite{Chen2019_YBC}.
We substituted the ATLAS 9 models for these stars instead. 

Based on this, we employed a slightly modified version of the assignment criteria in \cite{Chen2019_YBC}, namely:
\begin{enumerate}
    \item[(1)] Use ATLAS9 for stars with $T_\mathrm{eff} > 6500$ K.
    \item[(2)] Use PHOENIX BT-Settl for $T_\mathrm{eff} < 5500$ K.
    \item[(3)] Linearly interpolate between ATLAS9 and PHOENIX when $5500\,\mathrm{K} \leq T_\mathrm{eff} \leq 6500$ K.
    \item[(4)] Use COMARCS M/S star models when $(X_C /12)/(X_O /16) \leq 1$, $T_\mathrm{eff} < 4850$ K and $\log(g/\mathrm{cms}^{-2}) > 1.5$.
    \item[(5)] Linearly interpolate between PHOENIX and COMARCS M/S, where they overlap.
    \item[(6)] Use WM-basic models for $T_\mathrm{eff} > 20 000$ K and a mass loss rate $\dot{M} > 10^{-8} M_\odot \mathrm{yr}^{-1}$.
\end{enumerate}
Here, $X_C$ and $X_O$ refer to the surface mass fractions of $\mathrm{C}_{12}$ and $\mathrm{O}_{16}$, respectively. 

After selecting the correct spectral library for a given query source, we linearly interpolate the spectrum for the specific properties of the star, first in $T_\mathrm{eff}$, then in $\log(g)$ and finally in $Z$. 
The $T_\mathrm{eff}$-$\log(g)$ grids of the stellar atmosphere libraries are often not perfectly uniform, so that cases can occur where a query combination of ($T_\mathrm{eff}$, $\log(g)$) is not enveloped by enough spectra in the grid for an interpolation. 
In these cases, we fall back to assigning the spectrum of the nearest neighbour in the model grid to the query source. 
While this procedure could potentially introduce inconsistencies between the model stars and synthetic spectra, it appears to agree well with the output of the PARSEC web-interpolation tool, although \cite{Chen2019_YBC} does not explicitly state how such cases are handled in their code.

\subsubsection{Extinction law}
\label{sec:data_training_photometry_extinction_law}

Equation~\eqref{eq:AbsoluteMagnitude} indicates that we also need to specify the wavelength dependent extinction $A_\lambda$ in order to compute the magnitude of the star in the given filter. 
For this purpose, we selected the galactic extinction law as parametrised by \cite{Fitzpatrick1999}, which specifies $A_\lambda/A_V$ as a function of the slope of the reddening vector $R_V$. 
We selected this law, in particular, to facilitate the comparison to independent studies presented in Sect.~\ref{sec:results_obs_comparison}.
Given $R_V$ and $A_V$, we can then compute $A_\lambda$ and the attenuated flux $F_\lambda 10^{-0.4 A_\lambda}$ required in Eq.~\eqref{eq:AbsoluteMagnitude}.
We adopted a fixed $R_V = 3.1$ and sampled $A_V$ uniformly random between $0$ and $10$ during training of the cINN.
The extinction curve and its slope are known to vary across the Galaxy and in dense clouds \citep{Schlafly2016, Zhang2025}, and more recent parametrisations of the Milky Way extinction curve are available \citep{Fitzpatrick2019, Gordon2023}.
Because the photometry is computed on-the-fly, both the extinction law and $R_V$ could in principle be changed or treated as free parameters without regenerating any model grid, but this is outside the scope of this study. However, the effect of a mismatch in $R_V$ is briefly discussed in Sect.~\ref{sec:results_obs_comparison_starbystar}.

\subsubsection{Photometric system and zero points}
\label{sec:data_training_photometry_system}

For this study, we decided to compute photometric magnitudes for the Gaia DR3 \citep{Gaia2016, Gaia2023} and 2MASS \citep{Skrutskie2006A_2MASS} photometric filters in the VEGAMAG photometric system.
In the VEGAMAG system, we have $m_0 = 0$ and $f_\lambda^0$ corresponds to the spectrum of Vega ($\alpha$ Lyr). 
To be consistent with the PARSEC web-interpolator, we employed the CALSPEC database "alpha\_lyr\_stis\_010.fits" version\footnote{\url{https://ssb.stsci.edu/cdbs/calspec/}} of the Vega spectrum from \cite{Bohlin2020}. 
We note that this version of the Vega spectrum differs from e.g.~the one used in the official Gaia DR3 photometry \citep[they used "alpha\_lyr\_mod\_002.fits";][]{Riello2021}, which incurs minor zero-point offsets of $0.009$ mag in $G$, $0.015$ mag in $G_{BP}$ and $0.002$ mag in $G_{RP}$.
We did not use the official Gaia DR3 photometric zeropoints here, as \cite{Riello2021} explicitly states that they are incompatible with synthetic stellar atmospheres, such that zeropoints have to be recomputed using Eq.~\eqref{eq:AbsoluteMagnitude}. 

\subsection{Sampling and priors of the training set}
As we focus on extinction as our main extension of the \cite{Ksoll2020} approach in this study, we kept the metallicity fixed to solar metallicity ($Z_\odot=0.0152$ in the PARSEC framework) for our training set generation. 
An investigation of metallicity variation as an additional extension is planned for a follow up study.
For our training set, we employed a uniform random sampling in the age-initial mass space to be as unbiased as possible with regard to the evolutionary state of the stars. 
In this, we covered a range in \logage{} from $5$ to $10.1$ and from $0.09$ to $350$\,\msun{} in \mini.
Together with the adopted uniform sampling of $A_V$, these priors are deliberately agnostic and differ strongly from those of a Galactic field population, which is dominated by old, low-mass main-sequence stars with moderate extinction; in particular, about 40\% of the prior mass lies at ages below $10$\,Myr. 
Experiments with population priors based on Galaxy models \citep{Robin2003, Robin2012} and an initial mass function \citep{Kroupa2001} are out of the scope of this study and reserved to our follow-up work.

\begin{figure*}
    \centering
    \includegraphics[width=0.8\linewidth]{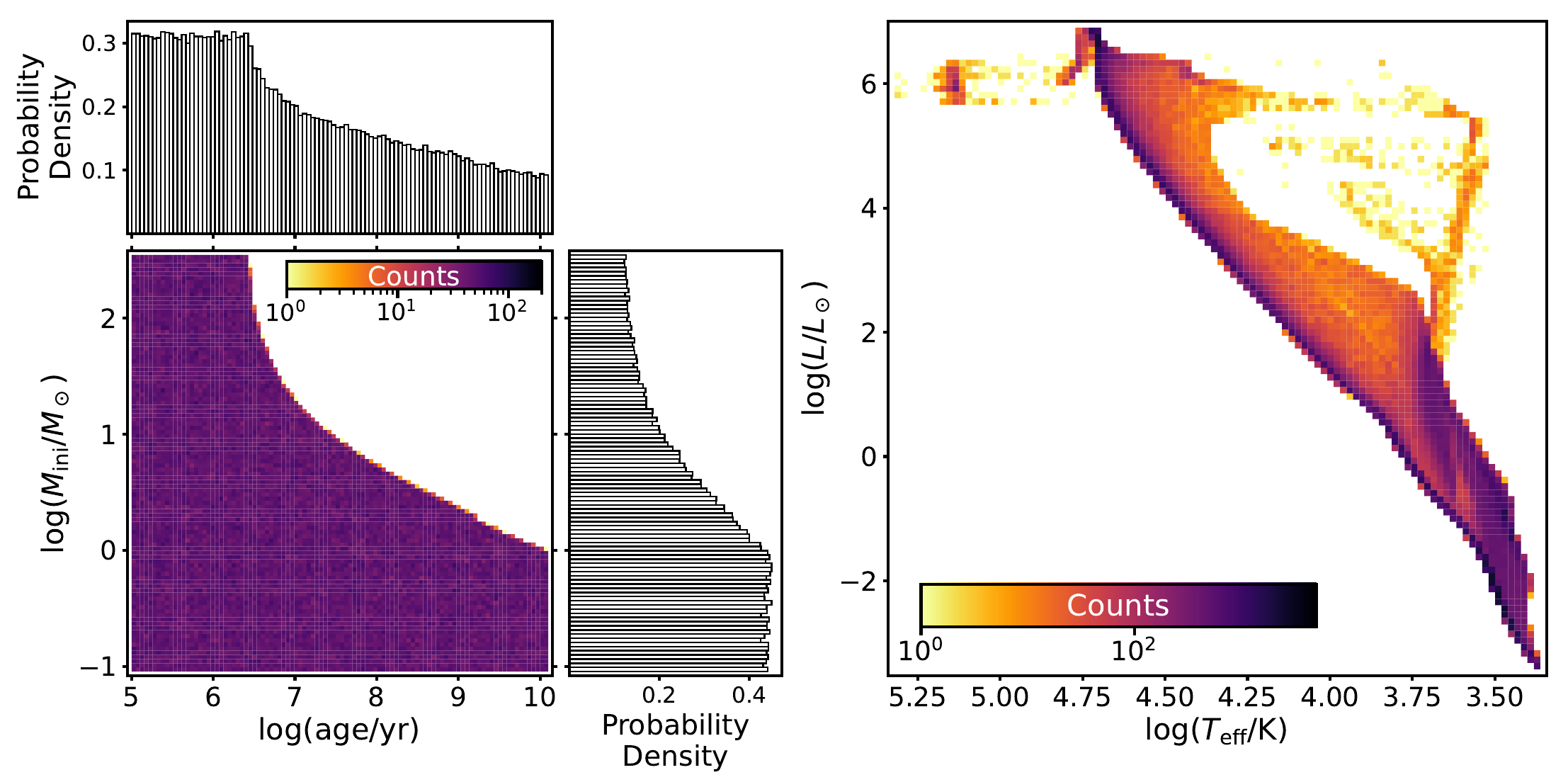}
    \includegraphics[width=0.8\linewidth]{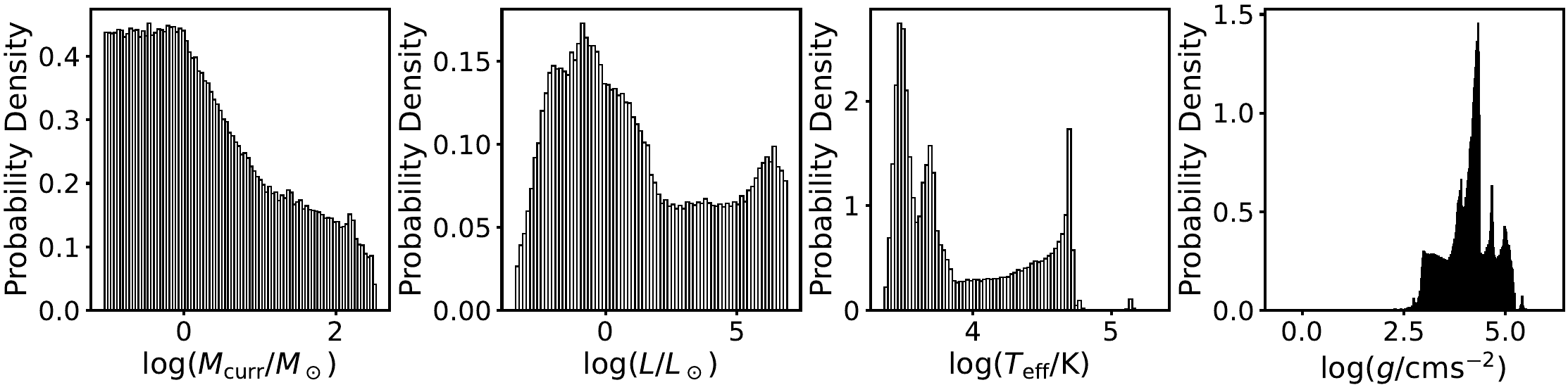}
    \caption{Prior distributions of target parameters in the generated training data. The top left panel shows a 2D histogram of the age vs initial mass priors, along with 1D histograms of the parameters individually. The top right panel displays the prior distribution of the training data in the Hertzsprung-Russell diagram. The four bottom panels indicate the 1D priors of current mass, luminosity, effective temperature and surface gravity respectively.}
    \label{fig:Sampling_TS_1}
\end{figure*}

In total, we generated $300\,000$ stellar models as the base for our training purposes. 
We further divided this initial database into dedicated training (${\sim}75\%$, $223\,200$ models), test (${\sim}20\%$, $61\,440$ models) and validation (${\sim}5\%$, $15\,360$ models) subsets. 
As previously described, the synthetic photometry for these models is then computed on-the-fly during training. 
Figure~\ref{fig:Sampling_TS_1} presents an overview of the resulting priors in \logage, \logmini, \logmcurr, \logl, \logteff and \logg, as well as the coverage achieved in the Hertzsprung-Russel diagram (HRD). 
As Fig.~\ref{fig:Sampling_TS_1} demonstrates,
our even sampling strategy in the age - initial mass space, leads to an uneven coverage in the HRD, where particularly some regions related to post main-sequence evolution are sparsely sampled (i.e.~in the upper right corner). The reason for this is that the models that populate this space lie on a very thin line in the \logage{} -- \logmini space, close to the locus that delineates the age at which stars of a certain initial mass "die" (i.e.~where the adopted PARSEC models would enter the TP-AGB phase and beyond). 
Populating the HRD evenly, would thus lead to a sampling in \logage{} -- \logmini that would be heavily biased towards the edge of this locus, as we have seen in \cite{Ksoll2020}.  

\section{Observational data}
\label{sec:data_obs}

\subsection{Target regions}
\label{sec:data_obs_sf_regions}
To gauge whether our approach is able to recover the highly differential extinction maps that are often found in regions of active star-formation \citep{Lombardi2011, Edenhofer2024}, we focused on two nearby star-forming regions for our tests on real data, namely Taurus and Upper Scorpius. 

\subsubsection{Taurus}
\label{sec:data_obs_sf_regions_taurus}
Taurus lies at a distance of about $140-150$ pc \citep{Zucker2020} and is located at the intersection of the Per-Tau shell and the Local Bubble \citep{Zucker2022, Soler2023}. 
It is one of the most extensively studied sites of nearby star formation \citep[see e.g.][for a review]{Kenyon2008} and host to a large population of less-massive pre-main sequence stars \citep{Soler2023}. 
The age of the young stellar objects in Taurus has been estimated to about $1-4$ Myr \citep{Liu2021, Luhman2023, Nayak2023}, while a secondary population of "field" and associated YSOs is believed to be between $8 - 56$ Myrs old \citep{Liu2021, Luhman2023}.
Gaia-based studies have characterised the three-dimensional structure and kinematics of the Taurus complex in detail \citep{Galli2019, Krolikowski2021}, and spectroscopic censuses provide spectral types and extinctions for its members \citep{Luhman2018, Esplin2019}. 
Its stellar population is organised in a hierarchy of groups, from ultra-wide pairs to dense nests \citep{Joncour2017, Joncour2018}, and the extinction of the cloud complex has been mapped with 2MASS colour-excess techniques \citep{Lombardi2010}.

\subsubsection{Upper Scorpius}
\label{sec:data_obs_sf_regions_usco}
Upper Scorpius is one of the youngest sub-groups of the Scorpius-Centaurus OB stellar association complex. 
At an estimated distance of $142-146$\,pc \citep{deZeeuw1999, Galli2018, Luhman2020}, Upper Scorpius is among the closest star-forming regions to our Sun and marks a prime test bed to investigate stellar feedback, as several studies suggest that the structure and star formation history of Sco-Cen has been heavily driven by feedback mechanisms \citep{Preibisch1999, MiretRoig2022, Posch2025, Armstrong2025}. 
The exact age of the Upper Scorpius group is still heavily debated, but has been estimated to be somewhere between $5-11$\,Myr \citep{DeGeus1989, Preibisch1999, Pecaut2016, David2019, MiretRoig2022, Armstrong2025}. 
Already with the first Gaia data release, \cite{Galli2018} derived a mean distance of $146 \pm 3 \pm 6$\,pc for the association and showed that its members are distributed in an elongated three-dimensional structure of about $74 \times 38 \times 32$\,pc for the brightest stars, with its major axis pointing roughly towards the Galactic centre and inclined by $20^\circ-25^\circ$ to the Galactic plane. 
The fainter members form a more elongated configuration (about $98 \times 24 \times 18$ pc), which the authors interpreted as an imprint of the star formation history of the region, the more massive stars having had more time to relax, in support of an age spread within the association.
Recent clustering studies based on Gaia astrometry revealed that Upper Scorpius is not a single stellar cluster, but exhibits a large degree of substructure in both kinematics and age \citep{Squicciarini2021, MiretRoig2022, BricenoMorales2023, Ratzenboeck2023}. 
This discovery prompted the proposition of a cluster-chain formation scenario, where older sub-clusters sequentially triggered the formation of the younger ones \citep{Posch2025}. 
However, this is still under debate, as the kinematic and temporal structure of Upper Scorpius may also be consistent with a scenario, where stellar feedback from Upper Centaurus-Lupus triggered star formation at different times and locations in Upper Scorpius \citep{Armstrong2025}.
A Gaia-based census with spectroscopic classifications and extinctions of its members is provided by \cite{Luhman2020}, and \cite{Lombardi2008} mapped the extinction of the adjacent Ophiuchus clouds has been mapped with 2MASS.

\subsection{Comparison catalogues}
\label{sec:data_obs_catalogues}
To ascertain the validity of the extinction maps derived with our extended cINN approach on real data, we decided to compare our results to independent extinction measurements from two studies that have analysed Gaia DR3 data. 
The first set of extinction estimates comes from the General Stellar Parametriser from Photometry (GSP-Phot) in the astrophysical parameters inference system \citep[APSIS;][]{Bailer-JonesEtAl2013_Apsis, CreeveyEtAl2023_Apsis} that is part of the official Gaia DR3 release.
GSP-Phot is a simulation-based Bayesian MCMC method that uses the low-resolution Gaia XP spectra, parallax and G-band magnitude to infer stellar properties and extinction. 
For a short review of GSP-Phot, see Appendix~\ref{app:gpspphot}.
We note that GSP-Phot employed the same \cite{Fitzpatrick1999} parametrisation of the extinction law, so that the monochromatic $A_0$ extinction value reported in Gaia DR3 corresponds to the same wavelength, $\lambda_0 = 541.4$\,nm, as the monochromatic $A_V$ estimated by our cINN approach.
We will, therefore, refer to this monochromatic extinction only as $A_V$ from here on. 

As our second comparison set, we chose the 3D extinction map derived by \cite{Zhang2025} and \citet{ZhangEtAl2023_Gaia_BayesModel_Data_APs}. 
They developed a machine learning approach to forward model Gaia XP spectra plus 2MASS and WISE photometry given stellar effective temperature, surface gravity, metallicity, parallax and extinction.
To train their approach, they used observational data for which the stellar properties have previously been spectroscopically estimated with the Large Sky Area Multi-Object Fibre Spectroscopic Telescope \citep[LAMOST;][]{Cui2012_LAMOST, Zhao2012_LAMOST}.
For inference, they embedded their ML forward model in a Bayesian framework to recover the above input parameters via maximisation of the stellar posteriors. 
In the more recent study, \cite{Zhang2025} further allowed for variations of the dust extinction law, estimating the slope $R_V$ of the reddening law in the optical as well as the magnitude of extinction.
For the purpose of comparing derived extinction maps in this study, we used the published catalogue from \cite{Zhang2025}.
This catalogue provides an extinction parameter and corresponding extinction curve for the conversion to the extinction at any given wavelength.
For our comparison, we converted this to the monochromatic extinction $A_V$ at $\lambda = 541.4$ nm to be consistent with GSP-Phot and our cINN.
We chose these two studies, in particular, because of their large coverage in the Gaia DR3 data and to have both a comparison to another simulation-based approach (GSP-Phot) and a purely observational data based method \citep{Zhang2025}.

At the same time, we want to emphasise that neither catalogue constitutes a ground truth, so that they serve only as independent points of comparison with known systematics.
GSP-Phot, for instance, is known to underestimate distances, and its metallicities are of limited reliability \citep{AndraeEtAl2023_GSP_Phot, Fouesneau2023, Babusiaux2023}.
Furthermore, both methods rely on the XP spectra \citep{DeAngeli2023, Montegriffo2023}, which restricts them to $G < 19$ for GSP-Phot and to the $G < 17.65$ sources with published XP spectra for \cite{Zhang2025}. 

\subsection{Gaia DR3 and 2MASS photometry}
\label{sec:data_obs_photometry}
Our observational datasets were compiled as follows.
For Taurus, we selected a window from $60^{\circ}$ to $75^{\circ}$ in right ascension (RA) and from $21.5^{\circ}$ to $30^{\circ}$ in declination (DEC), whereas for Upper Scorpius we extracted data between $235^{\circ}$ and $252^{\circ}$ in RA and $-29.5^{\circ}$ and $-16.9^{\circ}$ in DEC. 
To gauge the feasibility of recovering stellar parameters and extinction with our cINN approach from photometry alone, we only made use of the Gaia photometry in the G, \Gbp and \Grp bands of the sources in the selected regions. 
This delineates the first difference of our approach to GSP-Phot and \cite{Zhang2025}, which both relied on the low-resolution Gaia XP (RP/BP) spectra. 
We further complemented the Gaia photometry with infrared photometry from 2MASS in the J, H and Ks bands, as early tests with Gaia optical photometry alone returned highly uncertain extinction estimates. 
As such, we refined our data selection for Taurus and Upper Scorpius to require a crossmatch with 2MASS in Gaia DR3 \citep[using the Gaia best-neighbour cross-match;][]{Marrese2019}. 
This separates our coverage from GSP-Phot, as they only used optical Gaia data, but remains comparable to \cite{Zhang2025} as they complemented the Gaia XP spectra with infrared photometry from 2MASS and WISE. 

Requiring detections in both surveys determines the depth of our sample. 
The 2MASS point-source catalogue reaches $J \approx 15.8$, $H \approx15.1$ and $K_s \approx 14.3$ mag \citep{Skrutskie2006A_2MASS}, whereas Gaia reaches $G \approx 21$. 
For unreddened stars, the 2MASS limit therefore corresponds to $G \approx 16$ for solar-type stars and $G \simeq 17.5-18.5$ for M dwarfs, and is the binding constraint. 
Because extinction increases from $G$ to $K_s$ by about $0.7 A_V$, Gaia becomes the limiting survey only for strongly reddened stars, such as those in the cloud interiors.
The resulting apparent G-magnitude distributions of our samples are $18_{-1.8}^{+1.3}$ mag for Taurus and $17.6_{-1.8}^{+1.2}$ mag Upper Scorpius, with a minimum of $\sim21.1$ mag for both datasets.
We applied the standard corrections and quality cuts recommended for Gaia (E)DR3 photometry \citep{Riello2021, Fabricius2021}: the $G$-band magnitudes of sources with six-parameter astrometric solutions were corrected for their colour-dependent bias, and sources with an inconsistent corrected BP/RP flux excess factor $C^*$ were removed (i.e.~$|C^*| \geq 3\sigma_{C^*}(G) $), since their $G_\mathrm{BP}$ and $G_\mathrm{RP}$ photometry is affected by blending or by the background, which is particularly relevant in nebular regions and for faint sources. 
For 2MASS, we required ph\_qual = A or B in $J$, $H$, and $K_s$.

\subsection{Distances}
\label{sec:data_obs_distances}
As in \cite{Ksoll2020}, our approach takes the distance corrected photometry as an input (although now still subject to extinction). 
For this purpose, we cross-matched our Taurus and Upper Scorpius datasets with the geometric distance catalogue of \cite{Bailer-Jones2021} and used their median geometric distance ("r\_med\_geo") to infer the distance moduli, for which we corrected the input Gaia and 2MASS photometry.
This marks another difference in our analysis compared to the GSP-Phot results, which determines the spectrophotometric distance given the observed flux and astrometric parallax in a joint Bayesian framework, as well as the \cite{Zhang2025} approach, where parallax is self-consistently re-estimated.

We emphasize that the adopted distances already include the parallax zero-point correction of \cite{Lindegren2021}.
For sources with low parallax signal-to-noise ratio $\SNRPi \lesssim 5$, however, they are dominated by the Galactic prior of \cite{Bailer-Jones2021}, which does not use the photometry or the extinction of the star, so that the input distance modulus of these sources is partly a model assumption. 
A joint estimation of stellar parameters, extinction and distance within our framework is out of the scope here, but planned for our subsequent work.

\subsection{Final observational data sets}
\label{sec:data_obs_final}
Following the outlined data selection strategy, we ended up with initial samples of $415\,212$ stars for Taurus and $915\,804$ stars for Upper Scorpius. Out of these, there are $357\,255$ and $822\,546$ sources, respectively, with an extinction estimate ($A_0$) from GSP-Phot in Gaia DR3. \cite{Zhang2025} provide extinction measures for $234\,153$ sources out of the Taurus sample and for $649\,686$ objects in Upper Scorpius (with $\mathrm{quality\_flag} < 8$).
We note that for the analysis in the following we neglected both the uncertainties on distance, as well as the photometric errors, as we wanted to study the effects of adding extinction as a target parameter \citep[rather than being an input as in][]{Ksoll2020} on the predictive performance of our cINN approach in isolation.

\section{Method}
\label{sec:method}
\subsection{The conditional invertible neural network}
\label{sec:cinn_intro}
As in \cite{Ksoll2020} we employed a conditional invertible neural network \citep[cINN,][]{Ardizzone2019b} for this study, which is a variant of the NF that is tailored towards solving inverse problems. 
The latter refers to recovering the mapping $\mathbf{y} \rightarrow \mathbf{x}$ between the fundamental (physical) properties $\mathbf{x}$ of a system (e.g.~a star) and a set of corresponding observable quantities $\mathbf{y}$. 
This recovery is often challenging, because the associated forward process $\mathbf{x} \rightarrow \mathbf{y}$ that links system properties to the observables is frequently subject to an inherent loss of information, rendering the inverse problem degenerate. 

The cINN approach addresses these difficulties by introducing a latent space to encode this information loss. 
Specifically, a cINN learns a mapping $f$ from the physical parameters $\mathbf{x}$ to the latent variables $\mathbf{z}$, conditioned on the observables $\mathbf{y}$,
\begin{equation}
    \mathbf{z} = f(\mathbf{x}; \mathbf{c} = \mathbf{y}),
\end{equation}
while a prescribed prior distribution for the latent variables $P(\mathrm{z})$ is maintained during training.
It is usually a multivariate normal distribution with zero mean and unit covariance, 
This procedure encodes the variance of the target parameters that is not explained by the respective observables in the latent space.
The dimension of the latent space $\dim(\mathbf{z})$ is per construction of $f$ equal to the dimension of the target parameter space $\dim(\mathbf{x})$, whereas $\dim(\mathbf{y})$ can become arbitrarily large. 

Once trained, for a new observation $\mathbf{y}'$ the cINN can then sample from the prior of the latent variables to query the encoded information loss and, using its invertible architecture, generate samples of the posterior distribution $p(\mathbf{x}|\mathbf{y}')$ according to
\begin{equation}
    p(\mathbf{x}|\mathbf{y}') \sim g(\mathbf{z}; \mathbf{c} = \mathbf{y}') \,\,\,\mathrm{with}\,\,\, \mathbf{z} \propto P(\mathbf{z}),
\end{equation}
where $g(\cdot;\mathbf{c}) = f^{-1}(\cdot;\mathbf{c})$ denotes the inverse of the forward mapping, $f,$ for fixed condition, $\mathbf{c}$. 

A cINN consists of a series of invertible conditional affine coupling blocks \citep{Dinh2017, Ardizzone2019b}. To achieve invertibility these coupling blocks perform a set of complementary affine transformations on the halves $\mathbf{u}_1$ and $\mathbf{u}_2$ of their input vector $\mathbf{u}$, following
\begin{equation}
    \label{eq:coupling_forward}
    \begin{split}
        \mathbf{v}_1 &= \mathbf{u}_1 \odot \exp\left(s_2(\mathbf{u}_2, \mathbf{c})\right) \oplus t_2(\mathbf{u}_2, \mathbf{c}), \\
        \mathbf{v}_2 &= \mathbf{u}_2 \odot \exp\left(s_1(\mathbf{v}_1, \mathbf{c})\right) \oplus t_1(\mathbf{v}_1, \mathbf{c}), 
    \end{split}
\end{equation}
to compute the halves, $\mathbf{v}_1$ and $\mathbf{v}_2$, of the output vector $\mathbf{v}$. Here $\odot$ and $\oplus$ denote elementwise multiplication and addition, respectively. In Eq.~\eqref{eq:coupling_forward}, $s_i$ and $t_i$ represent arbitrarily complex transformations of the concatenation of $\mathbf{u}_i$/$\mathbf{v}_i$ and the conditioning input $\mathbf{c}$. To run the coupling block in reverse given the output vector $\mathbf{v} = (\mathbf{v_1}, \mathbf{v_2})$, Equation~\ref{eq:coupling_forward} is inverted according to
\begin{equation}
    \label{eq:coupling_backward}
    \begin{split}
        \mathbf{u}_2 &= \left(\mathbf{v}_2 \ominus t_1(\mathbf{v}_1, \mathbf{c})\right) \odot \exp\left(-s_1(\mathbf{v}_1, \mathbf{c})\right), \\
        \mathbf{u}_1 &= \left(\mathbf{v}_1 \ominus t_2(\mathbf{u}_2, \mathbf{c})\right) \odot \exp\left(-s_2(\mathbf{u}_2, \mathbf{c})\right),
    \end{split}
\end{equation}
where $\ominus$ denotes elementwise subtraction. As in both the forward (Eq.~\eqref{eq:coupling_forward}) and backward pass (Eq.~\eqref{eq:coupling_backward}) of the coupling block, the introduced transformations $s_i$ and $t_i$ are always only evaluated in the forward direction, $s_i$ and $t_i$ do not actually need to be invertible themselves. 
In practise, $s_i$ and $t_i$ are represented by small sub-networks (e.g.~a fully connected neural network) and their transformation is learned during the training of the cINN rather than prescribed beforehand \citep{Ardizzone2019a, Ardizzone2019b}. 

\subsection{Architecture and implementation details}
\label{sec:architecture_details}
We implemented the cINN approach in Python, using the dedicated Framework for Easily Invertible Architectures \citep[FrEIA\footnote{Available at \url{https://github.com/vislearn/FrEIA}},][]{Ardizzone2019a, Ardizzone2019b} package, which is based on the {\sc pytorch} \citep{Paszke2017_pytorch} deep learning module.
For the affine coupling blocks, we adapted the Generative Flow \citep[GLOW;][]{Kingma2018} architecture, which employs only two sub-networks per coupling block, such that $(s_1, t_1)$ and $(s_2, t_2)$ are jointly modelled by one sub-network each, respectively. 
Here we used simple, fully-connected sub-networks, comprised of three layers of size $512$ with rectified linear unit (ReLU) activation functions. 

The $\exp(s)$ term in Eqs.~\eqref{eq:coupling_forward} and \eqref{eq:coupling_backward} can potentially lead to an exploding magnitude problem, which may induce instabilities during training of the cINN.
To mitigate this issue, we adopted the clamping procedure of \citet{Ardizzone2019b}, which clamps the argument $s$ of the exponential function following
\begin{equation}
    s_{\mathrm{clamp}} = \frac{2\alpha}{\pi} \arctan\left(\frac{s}{\alpha}\right),
\end{equation}
where $\alpha = 1.9$.
Furthermore, we introduced random permutation layers before each affine coupling block, which permute the input vector of the coupling block in a random, but fixed (and thus invertible) manner,  
to better intermix the information between the two streams $\mathrm{u}_1$ and $\mathrm{u}_2$.
Our final cINN network architecture (as determined by hyperparameter optimisation, see Section~\ref{sec:training_strategy}) consisted of $14$ affine coupling blocks in the GLOW configuration.

\subsection{Inputs and target parameters}
\label{sec:InputsTargets}
As outlined in Sect.~\ref{sec:data_training_photometry}, our on-the-fly data augmentation scheme to compute synthetic photometry produces magnitudes for Gaia and 2MASS.
These six absolute magnitudes constitute our input feature space, that is $\mathbf{y} = \{M_\mathrm{G}, M_{\mathrm{G}_\mathrm{BP}}, M_{\mathrm{G}_\mathrm{RP}}, M_\mathrm{J}, M_\mathrm{H}, M_\mathrm{Ks}\}$.
In addition, we computed the integrated extinction $A_F$ in each of the six photometric bands (i.e.~$ F \in \{\mathrm{G}, \mathrm{G}_\mathrm{BP}, \mathrm{G}_\mathrm{RP}, \mathrm{J}, \mathrm{H}, \mathrm{Ks}\}$). 
We appended all six $A_F$ and $A_V$ to the target parameters, so that 
\begin{align*}
\mathbf{x} = 
\left\{
    \begin {aligned}
         & \log(\mathrm{age}), \log(M_\mathrm{ini}), \log(M_\mathrm{curr}), \log(L), \log(T_\mathrm{eff}), \log(g) \\
         & A_\mathrm{G}, A_{\mathrm{G}_\mathrm{BP}}, A_{\mathrm{G}_\mathrm{RP}}, A_\mathrm{J}, A_\mathrm{H}, A_\mathrm{Ks}, A_V
    \end{aligned}
\right\}.
\end{align*}
Although the extinction $A_F$ in the individual filters directly correlates with the monochromatic $A_V$ via the extinction law, we included them as target parameters because they are integrated quantities over each filter, so that they cannot be derived just from $A_V$, but actually depend on the spectrum of the star \citep{Jordi2010, Danielski2018, Casagrande2018}.
For narrow filters, $A_F$ could be approximated by $A_{\lambda_c}$ at the central wavelength $\lambda_c$ of the filter given $A_V$ and the extinction law, but for wide filters (e.g.\,Gaia G), discrepancies between $A_F$ and $A_{\lambda_c}$ can become non-negligible. 

\subsection{Additional data preprocessing}
As in \cite{Ksoll2020}, we linearly rescaled the data before ingesting them into the cINN.
Specifically, we applied the typical standardisation transformation to the target parameters $\mathbf{x}$, where each component is rescaled by their mean and standard deviation (i.e.~$\hat{x}_i = (x_i - \mu_{x_i}) /\sigma_{x_i}$).
For the input observations, on the other hand, we utilised a matrix whitening procedure \citep{Hyvarinen2000}, which produces a rescaled observable matrix (i.e.~the $223\,000\times 6$ matrix of all training synthetic observations) that has a unit covariance matrix. 
The coefficients for these linear rescaling operations are determined on the training data and then fixed during inference.
As the observables are computed on-the-fly during training, we retrieve the rescaling coefficients from one initial random realisation of the photometry for all $223\,000$ training data models.

\subsection{Training strategy and hyperparameter optimisation}
\label{sec:training_strategy}
To train the cINN, we followed \cite{Ardizzone2019b} and optimised the network weights $\Theta$ via minimisation of the maximum likelihood loss 
\begin{equation}
    \mathcal{L} = \mathbb{E}_i \left(\frac{||f(\mathbf{x}_i; \mathbf{c}_i, \Theta)||_2^2}{2} - J_i \right),
\end{equation}
where $J_i = \log\left|\det(\partial f/\partial \mathbf{x}|_{\mathbf{x} = \mathbf{x}_i})\right|$ represents the logarithm of the absolute value of the determinant of the Jacobian matrix evaluated at training example $\mathbf{x}_i$. 
In practise, we performed the weight optimisation using a stochastic gradient descent approach, namely the adaptive learning rate, momentum-based Adam \citep[adaptive moment,][]{Kingma2014} optimiser (with parameters $\beta_1 = \beta_2 = 0.8$ as determined by our hyperparameter optimisation).
The optimisation process also entailed the common L2 weight regularisation, with $\lambda = 2.34 \times 10^{-7}$, and we initialised Adam with a learning of $l_{\mathrm{init}} = 2.299 \times 10^{-3}$ (this corresponds to a scaling factor applied to the adaptive step size in the weight updates). 
In addition, we opted for a step-wise learning rate decay, applying a reduction by a factor of $\gamma = 0.643$ after every $5$ training epochs. 
We trained the cINN for a total of $200$ epochs, processing $72$ batches of size $1024$ in every epoch. 
We note that we did not loop over the entire base grid of our training data in every epoch for the purpose of reducing the overall training time. 
Although our photometry synthetisation scheme is fast enough to allow for on-the-fly data augmentation, doing so for $223\,200$ training examples in every epoch would still have notably slowed down training.
As our augmentation scheme, in essence, generates new and unique training examples in every epoch, not looping over the entire training set did not appear detrimental to the convergence of the cINN.
Training a single cINN according to this procedure took about 18.5h when making use of GPU acceleration on a NVIDIA RTX 2080Ti graphics card. 

We derived the set of hyperparameters outlined above (and in Sec.~\ref{sec:architecture_details}) using the Hyperband \citep{Li2018_Hyperband} algorithm for hyperparameter optimisation.
Hyperband combines adaptive resource allocation and early stopping to notably improve the efficiency of a random grid search for hyperparameters.
Given a maximum amount of epochs to train models for, Hyperband first constructs a set of differently sized brackets comprised of (randomly generated) hyperparameter configurations. 
Based on the size of the bracket, Hyperband then applies a more or less aggressive pruning strategy for the tested configurations. 
In the largest bracket, for instance, the network configurations are only trained for a few epochs at a time, before the least promising hyperparameter setups are pruned (based on the validation loss). 
This is repeated until only a single configuration is left in the bracket, which can then train for the prescribed maximum amount of epochs.
The interval between pruning steps increases as the bracket size decreases down to the smallest bracket, where a few configurations are allowed to train for the maximum number of epochs. 
This procedure strikes a balance between only training configurations until the end that appear promising early (i.e.~converge fast) and allowing for some slower converging models, which might converge to a better solution late in training. 
This allows Hyperband to explore a large set of hyperparameter configurations, while being much more time efficient than a normal randomised grid search. 
For more details, we refer to \cite{Li2018_Hyperband}.

\subsection{Posterior sampling, point estimates and sample rejection}
\label{sec:MAP_sample_rejection}

To evaluate the predictive performance of the cINN, it is helpful to derive point estimates based on the estimated full posterior distributions. 
For this purpose, we employed the maximum a posteriori (MAP) estimate, that is the most likely combination of stellar properties. 
To derive the MAP estimate, we used the change of variables formula to explicitly compute the estimated posterior probability density $p(\mathbf{x}_i|\mathbf{y})$ for each generated posterior sample $\mathbf{x}_i$, following
\begin{equation}
    \label{eq:change_of_variables}
    \begin{split}
    p(\mathbf{x}_i|\mathbf{y}) &= p_Z\left(f(\mathbf{x}_i;\mathbf{y})\right) \left|\det \frac{\partial f(\mathbf{x};\mathbf{y})}{\partial\mathbf{x}} \right|_{\mathbf{x}=\mathbf{x_i}} \\
    p(\mathbf{x}_i|\mathbf{y}) &= p_Z(\mathbf{z}_i) \left|\det \frac{\partial g(\mathbf{z};\mathbf{y})}{\partial\mathbf{z}} \right|_{\mathbf{z}=\mathbf{z_i}}^{-1},
    \end{split}
\end{equation}
where $f$ and $g$ denote the forward and backward mapping learned by the cINN, with $ f^{-1}(\cdot;\mathbf{c}) = g(\cdot;\mathbf{c})$. 
The MAP estimate then corresponds to the posterior sample with the highest posterior probability density $p(\mathbf{x}_i|\mathbf{y})$ in the full $13$ dimensional target parameter space. 
This guarantees a physically self-consistent combination of age, mass, luminosity, effective temperature, surface gravity and extinction, whereas the modes or medians of the individual marginal distributions, taken separately, need not necessarily correspond to a valid stellar model. 
In the posterior estimation step, we always generated 4096 posterior samples for each query star in our analysis. 
On an NVIDIA RTX 2080Ti GPU, this takes $297.9\pm1.5$\,s for a sample of $10\,000$ stars (averaged over 10 test runs) or $29.8\pm0.1$\,ms per star. 
This amortised inference cost is the main practical advantage of the cINN over MCMC-based approaches.
Generating 4096 posterior samples per query has proven to be a good trade-off between keeping inference times low and retrieving robust MAP estimates, as increasing the number of generated samples beyond this has not resulted in notable changes to the derived MAP values in our experiments. 
Nevertheless, this means that our MAP estimate technically only approximates the true maximum of the posterior density.
An even more precise value could be derived by combining the cINN with a stochastic gradient ascent approach, but this would incur a notably higher computational cost per query star.

In the posterior generation process, some of the generated samples may fall outside the parameter boundaries dictated by the training data. 
This occurs primarily in two scenarios. 
The first are edge cases, where the predicted posterior distributions may sometimes extend slightly beyond the training boundaries. 
The reason for this is that the cINN returns smooth posterior distributions and, thus, tends to struggle with adhering to hard boundaries. 
In the second scenario, the query falls outside of the parameter space learned by the network. 
If the query is only just outside, slight extrapolation is to be expected, such that some posterior samples may fall outside the training limits. 
If the query is vastly different from what the network has seen during training, however, it has to extrapolate a lot. 
In this case, there is no guarantee that the prediction is still sensible, as is typical for machine learning approaches \citep{Bishop2006_PatternRecognitionAndMachineLearning, Goodfellow2016_DeepLearning}. 
In order to avoid this potential issue, we introduced a sample rejection procedure in our analysis of the predicted posteriors, where we rejected all posterior samples that lie outside the formal target parameter limits of the training data (see Fig.~\ref{fig:Sampling_TS_1} for the limits). 
Accordingly, we also introduced a quality parameter $f_\mathrm{reject}$ for the predicted posteriors that keeps track of the fraction of posteriors samples (out of the initially generated $4096$ samples) that are rejected for each query input.

\section{Validation on synthetic data}
\label{sec:results_synthetic}

\begin{table*}
    \centering
    \caption{Summary of the performance metrics for the cINN MAP estimates for all target parameters.}
    \label{tab:synthetic_performance}
    \begin{tabular}{llccccccc}
    \toprule
    & \multicolumn{6}{c}{MAP estimates} & \multicolumn{2}{c}{Posteriors}\\
    \cmidrule(rl){2-7}\cmidrule(rl){8-9}
    Parameter & Units & MAE & RMSE & NRMSE & $\overline{\Delta}_\mathrm{MAP-GT}$ & $\sigma_\mathrm{MAP-GT}$ & MCE (\%) & MCE$_\mathrm{noSR}$ (\%)\\
    \midrule
    $\log(\mathrm{age}/\mathrm{yr})$ & dex & $0.159_{-0.155}^{+0.557}$ & $0.558$ & $0.109$ & $-0.113$ & $0.546$ & $4.10$ & $3.33$ \\
    $\log(M_\mathrm{ini}/M_\odot)$ & dex & $0.004_{-0.003}^{+0.031}$ & $0.106$ & $0.029$ & $0.017$ & $0.104$ & $2.87$ & $2.22$\\
    $\log(M_\mathrm{curr}/M_\odot)$ & dex & $0.004_{-0.003}^{+0.036}$ & $0.083$ & $0.023$ & $0.023$ & $0.080$ & $2.67$ & $2.16$ \\
    $\log(L/L_\odot)$ & dex & $0.003_{-0.003}^{+0.046}$ & $0.195$ & $0.019$ & $0.028$ & $0.193$ & $1.46$ & $1.54$\\
    $\log(T_\mathrm{eff}/\mathrm{K})$ & dex & $0.001_{-0.001}^{+0.017}$ & $0.070$ & $0.036$ & $0.011$ & $0.069$ & $2.36$ & $1.89$\\
    $\log(g/\mathrm{cms}^{-2})$ & dex & $0.007_{-0.006}^{+0.061}$ & $0.161$ & $0.024$ & $0.041$ & $0.156$ & $1.64$ & $1.19$ \\
    $A_{G_{BP}}$ & mag & $0.006_{-0.005}^{+0.032}$ & $0.172$ & $0.017$ & $0.035$ & $0.168$ & $1.17$ & $1.13$ \\
    $A_{G}$ & mag & $0.005_{-0.003}^{+0.026}$ & $0.137$ & $0.018$ & $0.028$ & $0.134$ & $1.28$ & $1.20$ \\
    $A_{G_{RP}}$ & mag & $0.004_{-0.003}^{+0.018}$ & $0.097$ & $0.017$ & $0.019$ & $0.095$ & $1.23$ & $1.16$ \\
    $A_{J}$ & mag & $0.002_{-0.001}^{+0.008}$ & $0.042$ & $0.016$ & $0.008$ & $0.041$ & $1.26$ & $1.20$\\
    $A_{H}$ & mag & $0.001_{-0.001}^{+0.005}$ & $0.027$ & $0.016$ & $0.005$ & $0.026$ & $1.24$ & $1.19$ \\
    $A_{Ks}$ & mag & $0.001_{-0.001}^{+0.003}$ & $0.018$ & $0.016$ & $0.004$ & $0.018$ & $1.24$ & $1.19$ \\
    $A_V$ & mag & $0.007_{-0.005}^{+0.029}$ & $0.160$ & $0.016$ & $0.032$ & $0.157$ & $1.16$ & $1.13$ \\
    \bottomrule
    \end{tabular}
\end{table*}

\begin{figure*}
    \centering
    \includegraphics[width=\textwidth]{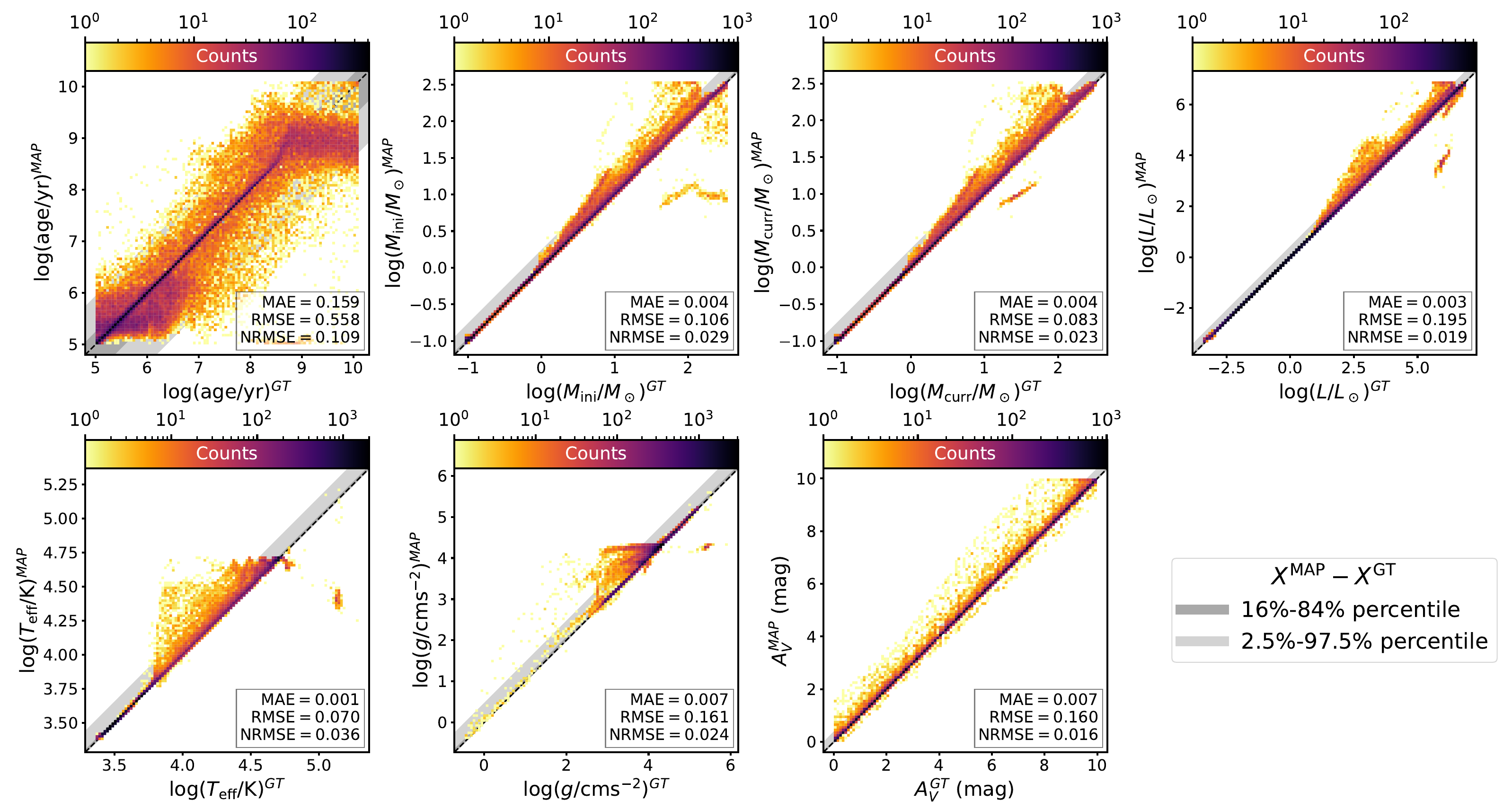}
    \caption{2D histograms, comparing cINN MAP estimates to ground truth values for 61440 synthetic test examples. The dark grey and grey shaded regions indicate the 68\% and 95\% confidence intervals of the residual $X^\mathrm{MAP} - X^\mathrm{GT}$, while the black dotted line corresponds to a perfect 1-to-1 mapping. We note that the colour-bar is logarithmically scaled in all panels for increased contrast.}
    \label{fig:MAP_vs_GT_1to1}
\end{figure*}

\subsection{Performance metrics}

After training the cINN, we evaluated its predictive performance on the $N= 61\,440$ held-out test stellar models. 
We generated the synthetic photometry for this test set in the same way as in the on-the-fly data augmentation step (see Sect.~\ref{sec:data_training_photometry}), sampling $A_V$ uniformly random between $0$ and $10$\,mag. 
To quantify the network performance, we calculated a series of measures based on the MAP point estimates for the target parameters. 
We computed the median absolute error (MAE) and its 16\%/84\% percentiles to gauge the average amplitude of the discrepancy between cINN MAP estimates and the ground truth properties. 
We also derived the root mean squared error (RMSE),
\begin{equation}
    \label{eq:rmse}
    \mathrm{RMSE} = \sqrt{\frac{1}{N}\sum_{i=1}^{N=61440} (X_{i, \mathrm{MAP}} - X_{i, \mathrm{GT}})^2},
\end{equation}
where $X_{i, \mathrm{MAP}}$ and $X_{i, \mathrm{GT}}$ denote the MAP estimate and ground truth for target parameter $X$, respectively, and the normalised RMSE (NRMSE),
\begin{equation}
    \label{eq:nrmse}
    \mathrm{NRMSE} = \frac{\mathrm{RMSE}}{\hat{X}_\mathrm{max} - \hat{X}_\mathrm{min}},
\end{equation}
where $\hat{X}_\mathrm{max}$ and $\hat{X}_\mathrm{min}$ refer to the upper and lower boundaries of target parameter $X$ in the training data. 
The NRMSE serves as a normalised measure that allows a direct comparison of the performance between the target parameters.
We also computed the commonly used mean $\overline{\Delta}_\mathrm{MAP-GT}$,
\begin{equation}
    \label{eq:bias}
    \overline{\Delta}_\mathrm{MAP-GT} (X) = \frac{1}{\mathrm{N}} \sum_{i=1}^{N} X_{i, \mathrm{MAP}} - X_{i, \mathrm{GT}},
\end{equation}
and the standard deviation $\sigma_\mathrm{MAP-GT}$,
\begin{equation}
    \label{eq:sigma}
    \sigma_\mathrm{MAP-GT}(X) = \sqrt{\frac{\sum_{i=1}^{N}\left| (X_{i, \mathrm{MAP}} - X_{i, \mathrm{GT}}) -\overline{\Delta}_\mathrm{cINN-GT}(X)\right|^2}{N-1}},
\end{equation}
of the residuals to gauge bias and average error of our point estimates, respectively.

Lastly, to summarise the overall quality of the predicted posterior distributions, we checked the posterior mean calibration errors for each target parameter.
This measure quantifies the average discrepancy between the nominal credibility intervals $q$ of the predicted posteriors and their actual empirical coverage $q_\mathrm{emp}$.
For a given credibility level $q$, the calibration error is defined as the difference
\begin{equation}
    \label{eq:calibration_error}
    e_\mathrm{cal}(q) = q_\mathrm{emp} - q = \frac{N_\mathrm{inliers}}{N} - q,
\end{equation}
where $N_\mathrm{inliers}$ denotes the number of test examples, where the ground truth falls into the $q$ credible interval, and N is the total size of the test data.
A negative $e_\mathrm{cal}$ indicates that the predicted posterior distributions are overconfident at the credibility level $q$ (i.e.~to narrow), whereas a positive value implies underconfidence (i.e.~too wide posteriors).
The mean (absolute) calibration error (MCE) is then derived as the average across all $K$ tested credibility levels, that is
\begin{equation}
    \label{eq:mce}
    \mathrm{MCE} = \frac{1}{K}\sum_i^{K}|e_\mathrm{cal}(q_i)|.
\end{equation}
Here, we computed the MCEs over a series of credibility intervals from $1\%$ to $99\%$ in steps $1\%$.

Table~\ref{tab:synthetic_performance} provides a summary of these performance metrics for all target parameters for the test on the held-out stellar models. 
For the posterior calibration, we provided a result for the predicted posteriors with and without sample rejection (as described in Sect.~\ref{sec:MAP_sample_rejection}). 
As Table~\ref{tab:synthetic_performance} demonstrates, the calibration of the predicted posteriors is overall quite excellent, falling in the range of $1.16$ to $4.1\%$ for all parameters.
Unsurprisingly, the artificial truncation introduced by our sample rejection procedure did slightly increase the MCEs, but the posteriors remained well calibrated regardless.
Comparing the NRMSEs, we found that the cINN MAP estimates appear to most accurately recover the extinction in all six filters as well as $A_V$ with NRMSE values of about $0.017$. 
Luminosity, \mini, \mcurr, surface gravity and effective temperature were also comparatively well constrained with NRMSEs between $0.019$ and $0.036$. 
The hardest parameter to constrain, relative to the other targets, was the age with an NRMSE of $0.109$, similar to our findings in \cite{Ksoll2020}.%, where we also identified age as the hardest parameter to predict.
Looking at the average magnitudes of the discrepancies between MAP estimates and ground truth in terms of the RMSE and $\sigma_\mathrm{MAP-GT}$, we found quite satisfactory results for most of the stellar properties. 
Specifically, the RMSEs/$\sigma_\mathrm{MAP-GT}$ for \logteff, \mcurr, \mini, \logg and \logl are $0.07$, $0.08$, $0.1$, $0.16$ and $0.19$ dex, respectively. 
For the predicted extinction, we have a wavelength dependent average discrepancy between $0.018$ mag for $A_\mathrm{Ks}$ to $0.17$ mag for $A_\mathrm{G_{BP}}$, which is consistent with the increase in the extinction range as the wavelength decreases.
For \logage, we found again the largest average discrepancies of the order of $0.55$\,dex in both RMSE and $\sigma_\mathrm{MAP-GT}$. 
In terms of the bias $\overline{\Delta}_\mathrm{MAP-GT}$ of the MAP estimates, Table~\ref{tab:synthetic_performance} shows that they are negligible for almost all target parameters, being on the order of up to only a few $0.01$.
The exception is once again \logage, where we identified a bias of about $-0.11$\,dex, indicating an average tendency in the MAP ages for underestimation.

\subsection{Accuracy of point estimates and outliers}

To illustrate the performance more in detail, Figure~\ref{fig:MAP_vs_GT_1to1} shows a direct one-to-one comparison of the cINN MAP estimates to the corresponding ground truth values for the synthetic test set. 
We note that we show only $A_V$ here as a representative value for the extinction estimates to keep the figure concise. 
As Fig.~\ref{fig:MAP_vs_GT_1to1} demonstrates, the cINN achieves a near one-to-one recovery of the ground truth for most parameters over the majority of the parameter range.
The main exception is again the stellar age, where we found larger discrepancies. %. and more scatter around a one-to-one recovery. 
Although we observed a very good agreement for most of the test data (we emphasise the logarithmic scaling of the colour bar in Fig.~\ref{fig:MAP_vs_GT_1to1}) between 5 and 8.5 in \logage, there is a notable amount of scatter around the one-to-one correlation between 5 and 7 in \logage. 
In addition, for the oldest stars, that is $\logage > 8.5$, the age appears to be systematically underestimated, as the network tends to return MAP estimates of about $\logage \approx 9$ for most stellar models. 
This tendency for underestimation for the very old stars likely drives the overall bias in the \logage estimates of $-0.11$ (see Table~\ref{tab:synthetic_performance}).
While the MAP predictions for the other target parameters showed notably less scatter around the one-to-one correlation than for \logage, the (few) outliers for \logmini, \logmcurr, \logl, \logteff and \logg tend towards an overestimation with the exception of very few stellar models. 
Looking at the most severe of these overestimated outliers (i.e.~those above the 97.5\% percentile of the residual distribution), we identified them as stellar models that fall primarily into areas of the HRD that are comparatively poorly sampled in our training data (see Fig.~\ref{fig:synth_hrd_outliers} in the appendix for an illustration of the outliers), such as the region towards the upper right of the main sequence or the small concentration of very massive stars in the top left corner of the HRD (compare to Fig.~\ref{fig:Sampling_TS_1}). 
These larger discrepancies could, therefore, at least in part be a result of the chosen training data sampling, which aimed to achieve an even sampling in the \logage-\logmini space, but not in the HRD. 
While a notable fraction of these outliers are stellar models in their post-main-sequence phase, there are also a few pre-main-sequence models among them, located in the region where post-main-sequence and pre-main-sequence overlap in the HRD (e.g.~close to the main-sequence turn-on or turn-off). 
In the latter cases, intrinsic degeneracies in the inverse problem can be an explanation for these discrepancies, as we have demonstrated in \cite{Ksoll2020}.
Regarding the few cases of severe underestimation in the MAP estimates for the target parameters besides age, we identified them as very hot stellar models that correspond to very massive stars ($>300\,\msun$) at the edge of the parameter space, which again are not as well sampled in our training data. 
For the age prediction, we identified a different behaviour with many outlier cases being located on the main sequence.  
This is not all too surprising, however, as the age of main sequence stars is notoriously hard to determine from photometry alone, as (especially for low-mass stars) the SEDs of these sources do not vary much for most of their main-sequence lifetime, rendering the age estimation highly degenerate for these stars \citep{Soderblom2010}.

\subsection{Self-consistency: resimulation}
To further verify the validity of the MAP estimates of our cINN, we recomputed the corresponding photometry with our full model interpolation and photometry synthetisation pipeline, using only the MAP estimates of \logage, \logmini and $A_V$ as input. 
For a summary of the resimulation results see Fig.~\ref{fig:resim_testset} in the appendix. 
Specifically, we found that the resimulated photometry is overall consistent with the input photometry to the network, closely tracing a 1-to-1 recovery with RMSEs between $0.17$ to $0.28$ mag across the six different filters.
This is, in particular, also the case for most of the stellar models with $\logage \geq 8.5$, for which we found the largest discrepancies between the true age and the MAP estimates. 
This directly demonstrates the degeneracy in age for these mostly low-mass main-sequence sources, where the photometry of these stars barely changes for a large fraction of their main sequence evolution. 
Nevertheless, there were also a few outliers in this resimulation experiment, where the resimulated photometry did not match the input photometry well, in particular in the 2MASS bands with discrepancies exceeding a few magnitudes.
For these cases, we identified them once again as stellar models that lie in the sparsely populated post-main-sequence region in the HRD, indicating that the cINN has indeed not learned to solve the inverse problem as well there as in the better sampled regions.

\section{Application to Taurus and Upper Scorpius}
\label{sec:results_taurus_usco}

\begin{figure*}
    \centering
    \includegraphics[width=0.9\linewidth]{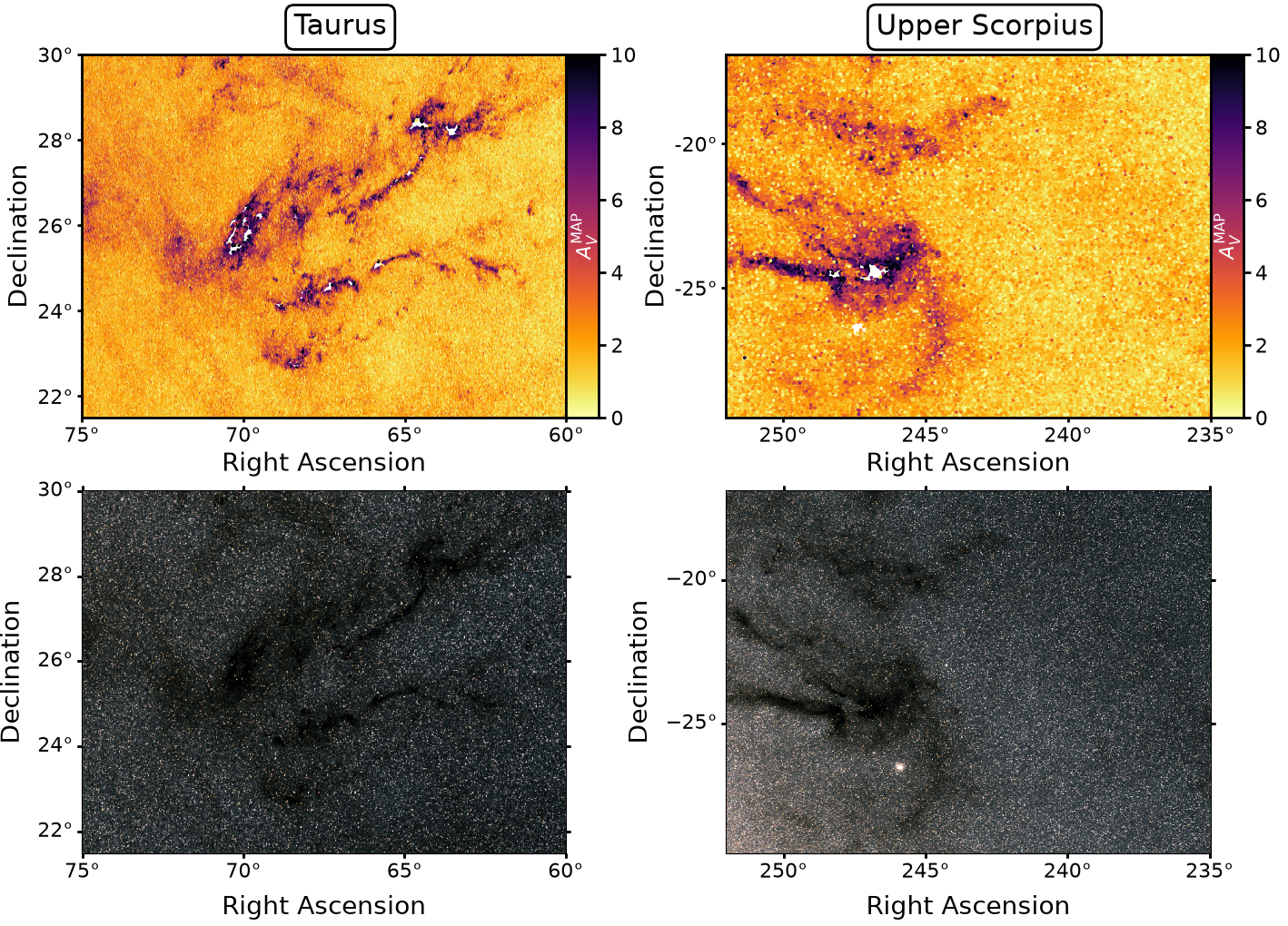}
    \caption{Extinction maps and colour images of the star-forming regions Taurus and Upper Scorpius. The extinction maps in the top panels are scatter plots, where each star is colour-coded according to the MAP extinction estimate from our cINN. The bottom panels present cutouts of the Gaia EDR3 "colour of the sky"-map (\url{https://sci.esa.int/s/WLy3Xp8}, image credit: ESA/Gaia/DPAC, A. Moitinho, license: CC BY-SA 3.0 IGO) for the fields of view of Taurus and Upper Scorpius. To align the Gaia EDR3 colour images with the extinction maps, we reprojected them from Galactic to equatorial coordinates. }
    \label{fig:cinn_exmap}
\end{figure*}

\begin{figure*}
    \centering
    \includegraphics[width=0.99\linewidth]{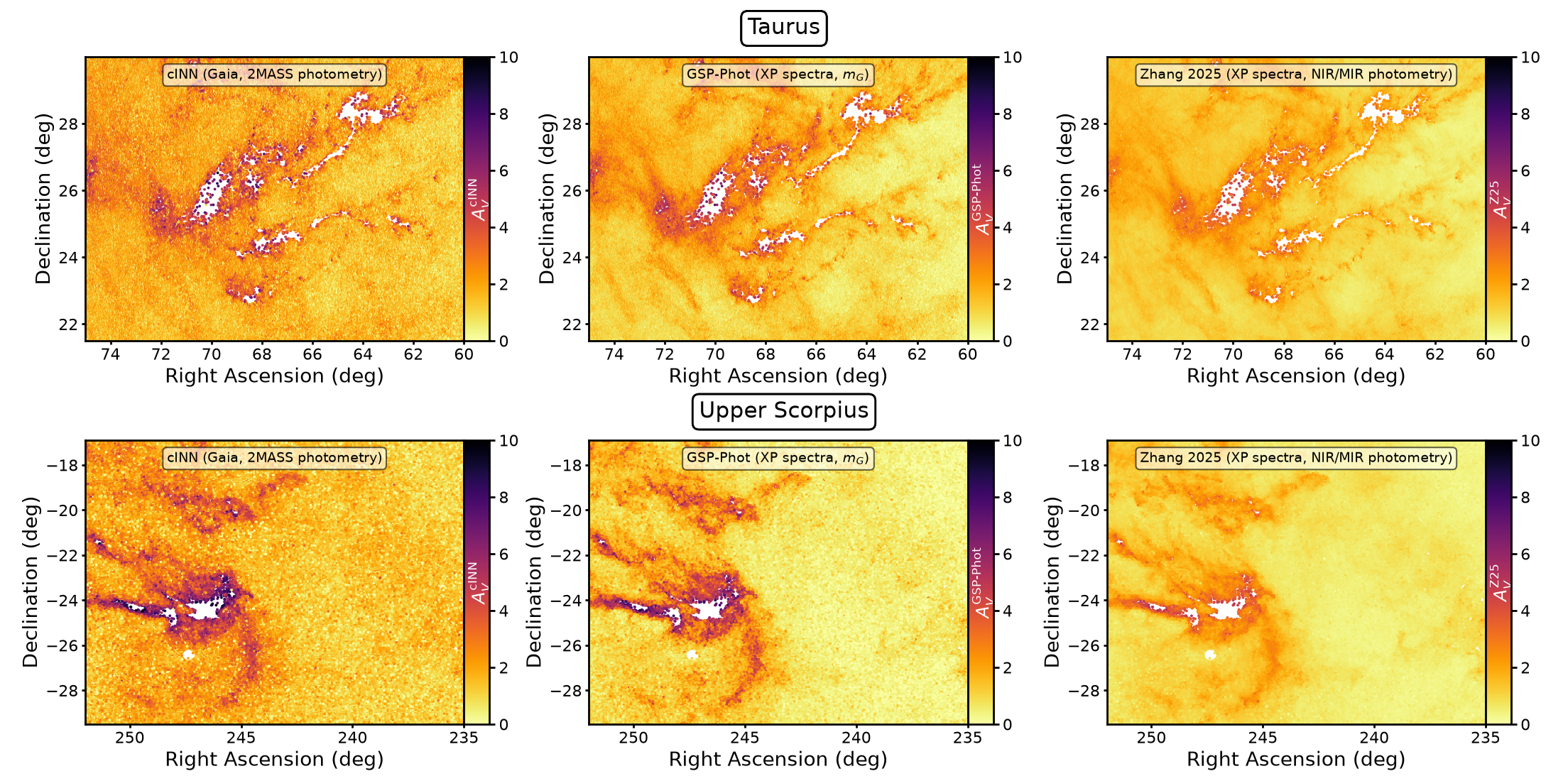}
    \caption{Comparison of the extinction maps for Taurus and Upper Scorpius between the cINN, GSP-Phot and the \cite{Zhang2025} approach. We note that this plot only shows stars for which a result exists from all three analysis methods. Each panel shows a scatter plot, colour-coded according to the respective extinction estimates.}
    \label{fig:cinn_exmap_vs_gspphot_z25}
\end{figure*}

After our tests on synthetic data established the feasibility of predicting extinction, we applied our trained cINN model to the Gaia and 2MASS data of Taurus and Upper Scorpius. 
The following will focus primarily on the results for the prediction of extinction, because neither GSP-Phot nor \cite{Zhang2025} provide estimates of stellar ages or masses. 
We reserve a more dedicated analysis of the joint recovery of stellar ages, masses and extinction to a follow-up study along with further improvements to our methodology.

\subsection{Valid predictions and rejected sources}
\label{sec:results_obs_sample_rej}

Applying our trained cINN to the observational datasets, we retrieved valid predictions for $360\,669$ sources in Taurus and $813\,086$ in Upper Scorpius.
For the remaining query stars, that is $54\,543$ ($13.1\%$ of the input data) in Taurus and $102\,718$ ($11.2\%$) in Upper Scorpius, our strict posterior sample rejection scheme, which did not allow for any extrapolation  (see Sect.~\ref{sec:MAP_sample_rejection}), has removed all generated posterior samples. 
Investigating these sources where $f_\mathrm{reject} = 1$, we found that the corresponding observations tended to have higher photometric uncertainties in the 2MASS bands (in particular in Ks) than the stars with valid results (i.e.~$f_\mathrm{reject} < 1$).
Specifically, we saw that the distributions of photometric errors (see Fig.~\ref{fig:sample_rejection_vs_phot_errs} in the appendix for a visualisation) for the population with no prediction (after posterior rejection) are quite broad and peak in the range of $0.1$ to $0.15$ mag in Ks, whereas the sample with valid predictions peaks distinctively at very small, i.e.~$<0.025$ mag, photometric uncertainties (although a long tail is present). 
As our network does not yet account for photometric uncertainties, it is not that surprising that input data with larger photometric uncertainties may push our method into a subspace of the input observational space, where it starts to extrapolate.  
A secondary experiment, where we trained a cINN only on Gaia photometry, provided further evidence that the 2MASS photometric uncertainties may be the primary cause of this issue.
Although this Gaia-only network struggles overall much more to constrain the extinction (much broader and uncertain posteriors for all target parameters, also on synthetic data), when applied to the real Gaia data, the fraction of sources with $f_\mathrm{reject}=1$ decreases notably (to just $1\%$ for the Taurus data for example).
As the Gaia photometric uncertainties are on average about a factor of 5-10 smaller than for 2MASS photometry, the effect of not accounting for the uncertainties is likely diminished in this experiment, resulting in fewer cases where the cINN extrapolates.   
This highlights the importance of properly accounting for photometric uncertainties, which will be subject of our follow-up development. 

\subsection{cINN extinction map}
\label{sec:results_obs_extinction_map}
Figure~\ref{fig:cinn_exmap} shows the resulting extinction maps for Taurus and Upper Scorpius based on the MAP estimates from our cINN. 
We note that these extinction maps are scatter plots that are colour-coded according $A_V^\mathrm{MAP}$ and do not include any additional spatial smoothing or aggregation.
Overall, our extinction map in Fig.~\ref{fig:cinn_exmap} presents a coherent picture, with higher extinction areas tracing the denser (optical dark) cloud structures in the two star-forming regions, whereas the fore- and background fields exhibit consistently lower extinction. 
Both fields of view show a similar distribution in extinction, although the extinction in Taurus is on average slightly higher with a median $A_V^\mathrm{MAP}$ of $1.98_{-1.04}^{+1.25}$\,mag than in Upper Scorpius with a median of $1.80_{-0.91}^{+1.29}$\,mag.
Our star-by-star approach also incurred a noticeable amount of noise across the extinction maps, where regions of low extinction belonging to the background fields are speckled with higher extinction solutions for individual stars.
Some degree of this noise appears intrinsic to our method, perhaps related to the limited photometric coverage used to jointly constrain extinction and the stellar properties.
It is worth emphasising, though, that at this stage we did deliberately not apply any post-processing to the extinction maps, and have in particular not made any use of the full posterior distribution function of extinction values.
The latter could for instance be utilised in a post-processing approach that takes the spatial correlation of extinction between neighbouring stars into account when deriving the point estimates from the predicted posterior distributions, rather than just selecting the MAP value.
On the other hand, we found that the degree of noise in our predicted extinction maps correlates with the signal to noise ratio of the Gaia parallax, \SNRPi.
Applying higher quality thresholds on \SNRPi (as is illustrated in Fig.~\ref{fig:cINN_extinction_map_vs_snr} in the appendix) notably decreases the degree of noise in our extinction map. 
Although we are not using the Gaia parallaxes directly in our distance correction but rather the geometric distances from \cite{Bailer-Jones2021}, the uncertainties of the latter are highly proportional to \SNRPi.
As the cINN is not yet accounting for these distance uncertainties in the predicted posterior distributions, it is therefore not surprising that the predictive performance may deteriorate for the data with low \SNRPi, producing noise in the otherwise low-extinction field regions.

\subsection{Resimulation}
\label{sec:results_obs_resimulation}
Analogous to our analysis on the synthetic data, we quantified the agreement between the photometry that corresponds to the cINN predictions of the stellar properties and the input Gaia/2MASS observations by re-running our full synthetic data pipeline given the MAP estimates of \logage, \mini and $A_V$. 
Although we found a considerable scatter in this experiment, the resimulated photometry traced a 1-to-1 recovery on average, with RMSEs ranging from about $0.7-0.8$ mag in the Ks band to $1.4$ mag in $G_\mathrm{BP}$ (for a visualisation see Fig.~\ref{fig:resim_taurus_usco} in the Appendix).
As our cINN did not account for photometric or distance uncertainties during the prediction, some scatter is naturally to be expected here. 
The hypothesis that distance uncertainties are a driving factor of the discrepancies in this resimulation experiments is further supported when looking at the subset of data with low distance uncertainties, i.e. $\SNRPi > 20$.
For this subset, the scatter around a perfect 1-to-1 recovery reduced drastically with RMSEs decreasing to about $0.45$ mag in the Ks-band and $0.6-0.7$ in $G_\mathrm{BP}$ for both the Taurus and Upper Scorpius datasets.
Although these residuals remain notably larger than the photometric errors (see Fig.~\ref{fig:sample_rejection_vs_phot_errs}), this result indicates that the cINN predictions are overall reasonably consistent with the input photometry, in particular, as we cannot necessarily expect a network trained without noise to perfectly fit noisy photometry within the observational errors.
Nevertheless, this further highlights the necessity of adapting noise-aware training in our subsequent development.

\subsection{Comparison with GSP-Phot and Zhang \& Green 2025}
\label{sec:results_obs_comparison}
To compare our extinction map to GSP-Phot and \cite{Zhang2025}, we focused in the following only on the subsets of data, for which all three analysis pipelines provided a result for the extinction.
This notably limited the sizes of the datasets, namely to $204\,313$ sources for Taurus and $600\,043$ sources for Upper Scorpius. 

\subsubsection{Extinction maps}
\label{sec:results_obs_comparison_exmaps}
Figure~\ref{fig:cinn_exmap_vs_gspphot_z25} shows a side-by-side comparison of the extinction maps derived for Taurus and Upper Scorpius based on our cINN, GSP-Phot and \cite{Zhang2025}.
In general, this comparison revealed a qualitatively good agreement with both GSP-Phot and \cite{Zhang2025}, where all three extinction maps exhibited similar high-extinction structures that spatially correlate to the (optically dark) clouds in the respective star-forming regions. 
The primary differences in the cINN extinction map compared to the GSP-Phot or \cite{Zhang2025} outcomes is a higher degree of noise and an on average higher predicted extinction, in particular in the regions of the background field.
The cINN found a median extinction of $1.79_{-0.86}^{+0.86}$\,mag for Taurus and $1.67_{-0.81}^{+0.94}$\,mag for Upper Scorpius, respectively, whereas GSP-Phot arrived at $1.26_{-0.59}^{+0.77}$\,mag and $0.85_{-0.53}^{+0.81}$\,mag, and \cite{Zhang2025} returned $1.18_{-0.41}^{+0.50}$\,mag and $0.83_{-0.34}^{+0.53}$\,mag.
While the cINN and GSP-Phot reached somewhat similar peaks in extinction ($9.9$ and $8.0$ mag for Taurus and $10$ and $9.4$ mag for Upper Scorpius), \cite{Zhang2025} remained more conservative, going up to only $5.4$ and $5.7$ mag for Taurus and Upper Scorpius, respectively.

\begin{figure}
\centering
    \includegraphics[width=\linewidth]{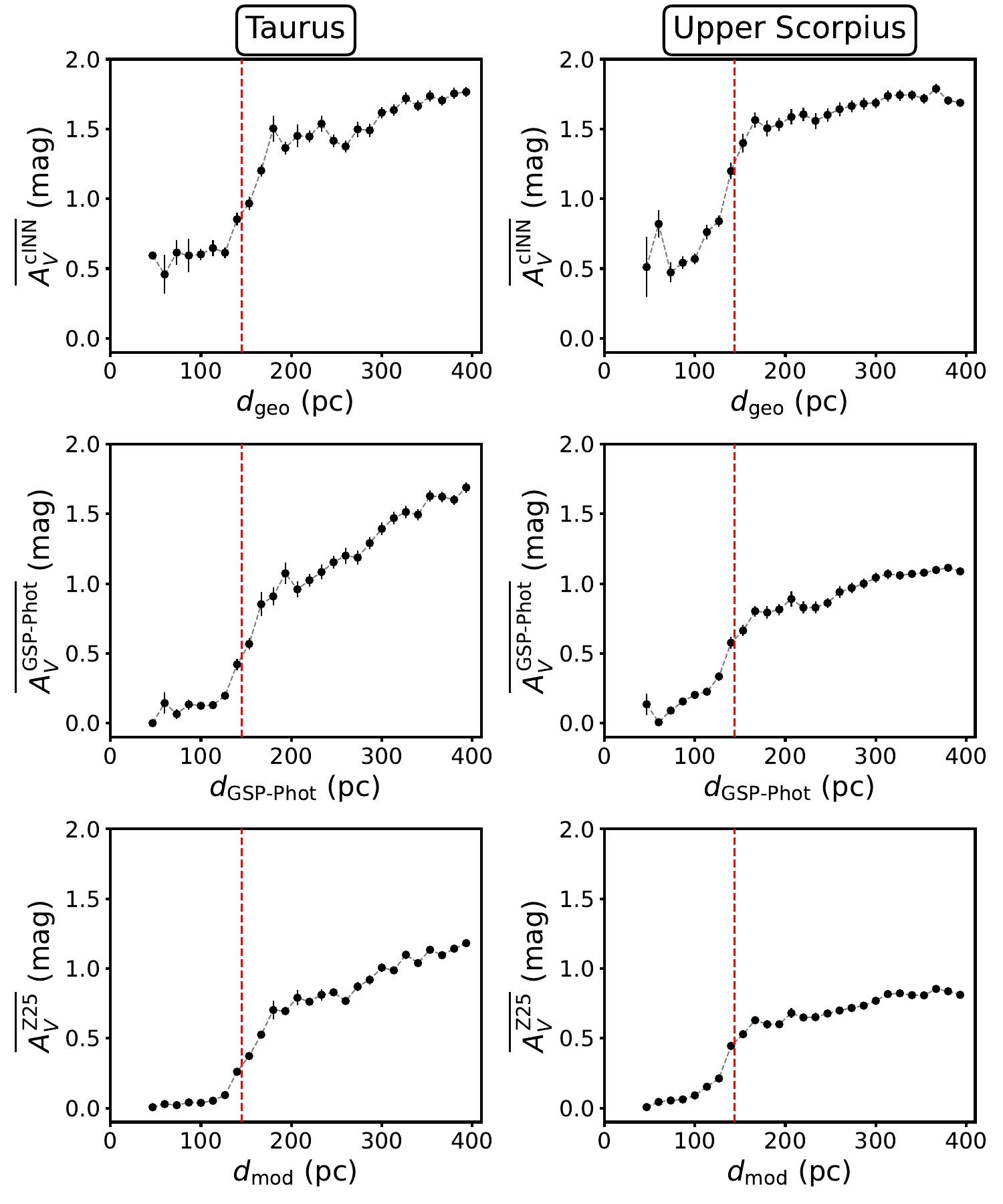}
    \caption{Comparison of the binned mean extinction as a function of stellar distance out to $400$ pc for Taurus and Upper Scorpius between cINN, GSP-Phot and \cite{Zhang2025}. We note that for the purposes of self-consistency, each method is plotted against the distance metric used in the respective analysis. This means that between the rows there might not be the exact same stars in each distance bin, but we ensured that all sources come only from the subset of data, where all three methods provided an extinction estimate. In all panels, the red dashed line marks the nominal distance of the Taurus and Upper Scorpius molecular clouds, respectively.}
    \label{fig:extinction_vs_distance_400pc}
\end{figure}

\subsubsection{Extinction versus distance}
\label{sec:results_obs_comparison_distVex}
As a secondary qualitative check of the structure of our predicted extinction map, we compared the average extinction as a function of distance in the selected fields of view between the cINN, GSP-Phot and \cite{Zhang2025}.
This type of analysis can for instance be used to derive distances to the star-forming regions in question \citep[see e.g.~][]{Zucker2020}.
Figure~\ref{fig:extinction_vs_distance_400pc} presents the outcome of this experiment.
We note that we selected here only stars out to a distance of $400$\,pc, in order to focus on the locations of the clouds in Taurus and Upper Scorpius.
Furthermore, we decided to plot the average extinction (with their standard deviation of the mean) as a function of the respective distance measure used in each analysis approach to maintain self-consistency. 
This means that for the cINN, we used the geometric distances from \cite{Bailer-Jones2021}, for GSP-Phot their own predicted distances ("distance\_gspphot" in Gaia DR3) and for \cite{Zhang2025}, we derived a distance $d_\mathrm{mod}$ from the modified parallax that they estimate jointly with the stellar parameters. 
We note that the binning was applied based on the respective distance measures, so that there are not necessarily the same stars in each distance bin between the three methods. 
Figure~\ref{fig:extinction_vs_distance_400pc} demonstrates again a good qualitative agreement between our cINN predictions and the GSP-Phot and \cite{Zhang2025} results, with all three methods exhibiting a similar structure in the average extinction.
In particular, all three outcomes show a notable jump in the average extinction at a distance that corresponds to the nominal distances of the respective star-forming regions (we assumed a distance of $145$ pc \citep{Zucker2020} and $144$ pc \citep{deZeeuw1999, Galli2018, Luhman2020} for Taurus and Upper Scorpius here).
However, Fig.~\ref{fig:extinction_vs_distance_400pc} also highlights the offset in extinction of about $0.5-0.7$ mag on average in our predictions, which extends even to sources in front of the clouds, where both GSP-Phot and \cite{Zhang2025} find average extinction close to zero.

\subsubsection{Star-by-star comparison}
\label{sec:results_obs_comparison_starbystar}
A direct star-by-star comparison revealed mean discrepancies of $0.46\pm0.77$ mag and $0.73\pm0.78$ mag for Taurus and Upper Scorpius with respect to the GSP-Phot results, whereas compared to \cite{Zhang2025} we found $0.59\pm0.71$ mag and $0.83\pm0.71$ mag, respectively (see Fig.~\ref{fig:av_map_vs_gspphot_z25} in the Appendix for a visualisation).
For GSP-Phot the discrepancies appeared as a flat offset to a 1-to-1 match, whereas in relation to \cite{Zhang2025} we also identified a notable slope. 
This difference in behaviour is likely explained by the fact that GSP-Phot uses the same extinction law as our cINN approach, namely \cite{Fitzpatrick1999} with a fixed $R_V = 3.1$, whereas in the \cite{Zhang2025} analysis the slope of the reddening law, $R_V$, is actually a free parameter that is jointly estimated with the extinction itself.
In particular, \cite{Zhang2025} found a considerable spread of $R_V$ across both fields of view with median $R_V$ of $2.5_{-0.2}^{+0.2}$ for Taurus and $3.0_{-0.2}^{+0.3}$ for Upper Scorpius, respectively. 
With such differences in $R_V$ some systematic offsets in the predicted extinction are to be expected, in particular in the case of Taurus, where their average $R_V$ is considerably lower than the fixed $R_V = 3.1$ that the network is trained on.
However, a preliminary prediction test on synthetic data with photometry generated with $R_V=2.5$ remained inconclusive.
Although we found a systematic deviation that manifests in a slope in a 1-to-1 comparison, that slope did not match the direction found in the comparison to \cite{Zhang2025}.
Specifically, we discovered that the cINN had a tendency towards underestimating the extinction on the synthetic test data generated with $R_V=2.5$ rather than reproducing the overestimation in Fig~\ref{fig:av_map_vs_gspphot_z25}. 
As methodological differences and the effects that are currently unaccounted for in our approach could potentially compound in non-obvious ways to induce these discrepancies between the methods, we decided to leave further in-depth studies of this aspect to our subsequent development at this stage.

\subsubsection{Luminosity - extinction degeneracy and distance}
\label{sec:results_obs_comparison_logl_vs_deltaAv}
To investigate the origin of the flat offset in our extinction estimates compared to the GSP-Phot results, we compared the predicted luminosities.
By default GSP-Phot does not return the stellar luminosity in the Gaia DR3 release, but it provides estimates of both stellar radius $R_*$ and effective temperature \teff. 
For this comparison, we, therefore, computed the corresponding GSP-Phot luminosities as
\begin{equation}
    L_*^\mathrm{GSP\text{-}Phot} = 4\pi R_*^2 \sigma \teff^4.
\end{equation}
Here, we identified an average discrepancy in \logl between the cINN and GSP-Phot estimates of $0.72\pm0.67$ dex and $0.61\pm0.69$ for Taurus and Upper Scorpius.
Limited to the highest quality data, i.e.\,$\SNRPi > 20$, the average discrepancy decreased to about $0.1\pm 0.2$ dex for both datasets (for a visualisation, see Fig.~\ref{fig:delta_av_vs_delta_logL_cinn_vs_gspphot} in the Appendix).
From the differences in luminosities $\Delta\logl$, we derived the absolute difference in the bolometric luminosity $\Delta M_\mathrm{bol}$ between the cINN and GSP-Phot estimates following
\begin{equation}
    \begin{split}
    \Delta M_\mathrm{bol} & = \left| -2.5 \log(L^\mathrm{cINN}/L^\mathrm{GSP\text{-}Phot}) \right| \\
    & = \left|-2.5 \left(\logl^\mathrm{cINN} - \logl^\mathrm{GSP\text{-}Phot}\right) \right|\\
    & = 2.5 \Delta\logl,
    \end{split}
\end{equation}
and compared this to the difference in the extinction estimate.
This analysis revealed a partial correlation between the discrepancy in the predicted extinction and the difference in estimated luminosity / bolometric magnitude, in particular, when we focus on the high-quality data ($\SNRPi > 20$, see bottom row of Fig.~\ref{fig:delta_av_vs_delta_logL_cinn_vs_gspphot} in the Appendix).
This suggests that, compared to GSP-Phot, our cINN tended to slightly overestimate the stellar luminosity, which it then compensated by an increase in the predicted extinction. 
Considering that our resimulation experiment demonstrated that the photometry that corresponds to the cINN predictions adheres quite well to the input Gaia/2MASS observations (in particular for the data with low distance uncertainty), part of these discrepancies in luminosity and extinction may not necessarily be a bias in our approach, but could also be explained by the inherent luminosity-extinction degeneracy of this inverse problem.  
In this context, it is worth noting that GSP-Phot does not employ a flat prior on extinction, contrary to our cINN approach, but implements a bias towards lower extinction solutions \cite{AndraeEtAl2023_GSP_Phot}. 
It is, therefore, possible that our more agnostic, flat prior may induce a preference for valid, higher extinction solutions, which are down-weighted in the GSP-Phot approach, resulting in the observed offset. 
A preliminary experiment, in which we modified the sampling of $A_V$ in our on-the-fly photometry synthetisation to use a Gamma function with a peak at 0 (keeping the network architecture and hyperparameters otherwise fixed), returned no improvement in the observed discrepancies, but slightly amplified them instead. 
This provides some evidence that the choice of prior might not be a driving factor in the observed discrepancies, but as we cannot rule out sub-optimal convergence of the network with this modified training strategy at the moment, we deem these results inconclusive for now and reserve further experimentation for a follow-up.

\subsection{Sources characterised only by the cINN}
\label{sec:results_obs_comparison_cinn_unique}
As a final part of our analysis, we briefly highlight the subset of Gaia data, for which only our cINN was able to return an extinction estimate, whereas neither GSP-Phot nor \cite{Zhang2025} have. 
We note that some of the "missing predictions" in either GSP-Phot or \cite{Zhang2025} may, of course, be a result of the applied quality cuts and selection criteria.
For our Taurus dataset, this comprised a total of $33\,993$ sources and for Upper Scorpius, we found $42\,435$ such cases, where only the cINN provided a result in our analysis.
In turn, we also point out that out of the $54\,543$ stars in Taurus and $102\,718$ stars in Upper Scorpius, where our analysis did not return a prediction after posterior sample rejection, for $49\,397$ and $97\,867$, respectively, there exists a solution from either GSP-Phot or \cite{Zhang2025}.
Looking at the cases with a result only from our cINN, we identified large fractions of them to be "lower" quality data with large uncertainties in both distance and photometry (e.g.~more than 50\% of these cases have $\SNRPi < 5$). 
Consequently, the corresponding extinction maps of these "cINN-unique" solutions tend to be quite noisy (see Fig.~\ref{fig:extinction_map_cinn_unique_results} in the Appendix for a visualisation).
Given our conclusions regarding the effects of uncertainties on our predictions so far, most of these "cINN-unique" results should be treated with caution.
However, focusing on the high-quality data among them (i.e.~$\SNRPi > 20$), we found some interesting structures, in particular, in the Upper Scorpius field of view.
A significant number of the "cINN-unique" outcomes were clustered within the optical dark clouds in these star-forming regions and appeared to probe the very high-extinction inner parts of these clouds.
At the same time, they exhibited a quite coherent inwards gradient in extinction (see in particular the bottom right panel in Appendix, Fig.~\ref{fig:extinction_map_cinn_unique_results}).
Although we could not identify a conclusive reason from the available Gaia DR3 flags as to why these particular observations do not have a solution in GSP-Phot, this may indicate an advantage of our photometry-only approach when probing the higher-extinction, denser cloud regions, where the Gaia XP spectra used by both GSP-Phot and \cite{Zhang2025} may potentially have been insufficient.
As our sample is limited by the 2MASS depth for most stars, however, the cINN is not necessarily expected to reach fainter magnitudes than GSP-Phot ($G < 19$) for typical field stars, so that the absence of GSP-Phot solutions for these sources are more likely related to the adopted quality filters than actual depth.
With respect to \cite{Zhang2025}, which is limited to the published XP spectra ($G < 17.65$), on the other hand, the cINN can reach fainter magnitudes for red and reddened stars, which are precisely those found in the cloud regions.

\section{Discussion and outlook}
\label{sec:discussion}

Overall, our results have demonstrated that our photometry-based cINN is able to qualitatively recover the structure of the extinction maps of GSP-Phot and \cite{Zhang2025} in both Taurus and Upper Scorpius.
However, our analysis showed that quantitatively our cINN predictions appear to systematically overestimate extinction in both star-forming regions compared to the other two studies.
We have already discussed some aspects that could potentially explain these discrepancies, including the photometric and distance uncertainties that are currently unaccounted for in our approach, differences in the choice of the priors on extinction, or the fact that \cite{Zhang2025} allows for variations of $R_V$ in their analysis.
In the following, we want to expand upon the effects that may affect the agreement between the three methods.

\subsection{Current limitations}
\label{sec:discussion_limitations}

\paragraph{Metallicity.}

In this study, we aimed to investigate the influence of predicting extinction with our approach rather than requiring it as an input \citep[as previously in ][]{Ksoll2020} in isolation, and, therefore, kept the metallicity fixed to $Z_\odot$ for our training data.
However, both GSP-Phot and \cite{Zhang2025} treated metallicity as a free parameter that is estimated along with the other stellar parameters.
Such discrepancies in metallicity are likely to affect the prediction of extinction as well.
Although the assumption of solar metallicity is certainly reasonable for e.g.~the young cluster members of Taurus and Upper Scorpius, the considered fields of view are dominated by the background field stars. 
In fact, both GSP-Phot and \cite{Zhang2025} identified a notable spread in metallicity across our datasets.
Focusing on the subsets, where all three methods returned an extinction estimate, GSP-Phot returns a metallicity range (in [M/H]) of $-4.13$ to $0.8$ for both star-forming regions with a median metallicity of $-0.79_{-0.79}^{+0.58}$ for the Taurus and $-0.68_{-0.71}^{+0.53}$ for the Upper Scorpius data.
Likewise \cite{Zhang2025} found predominantly sub-solar metallicity covering a total range in [Fe/H] of $-2.87$ to $0.76$ with medians of $-0.41_{-0.53}^{+0.32}$ and $-0.32_{-0.42}^{+0.36}$ for Taurus and Upper Scorpius, respectively.
We emphasise that these statistics represent the entire selected fields of view and not the cluster members of Taurus and Upper Scorpius.
A preliminary experiment with synthetic data generated for sub-solar metallicity confirmed that a metallicity discrepancy with respect to the training data results in a greater dispersion (around the ground truth) of the predicted extinction. 
The effect was comparably small (at least down to $0.1 Z_\odot$), however, and did not result in significant systematic offsets. 
This hints at metallicity being a contributing factor to the discrepancies we observed, albeit likely not a dominant one in the systematic offsets. 

\paragraph{Input information.}
We deliberately set out to built a tool that relies only on photometric observations to gauge its applicability and performance for survey data that may not include spectroscopic information. 
As discussed in Sect~\ref{sec:data_obs}, both GSP-Phot and \cite{Zhang2025} used Gaia XP spectra, however, which provide flux measurements at 61 wavelengths. 
GSP-Phot further combined these with the Gaia G-band magnitude, whereas \cite{Zhang2025} also added photometric information in the infrared up to $4.6\,\mu\mathrm{m}$. 
Although our approach complements the optical Gaia photometry with infrared 2MASS observations (compared to the purely Gaia-centred approach of GSP-Phot), this means that the cINN operates on considerably less information.
It is not implausible that such differences in the input observational space could result in systematic discrepancies in the inference of the stellar properties or the interstellar extinction.
Whether the inclusion of the XP spectra in our approach would further align our extinction predictions with GSP-Phot or \cite{Zhang2025}, is outside the scope of this work, however, as our synthetic data generation pipeline does not yet have the capability to emulate Gaia XP spectra.
We reserve such an experiment for our subsequent development of our approach.
In this context, we want to emphasise that the purpose of this study was not to produce "better" extinction maps than GSP-Phot or \cite{Zhang2025} in the first place, but to demonstrate that our photometry-only approach can reasonably reproduce results that are based on substantially more observational constraints, such as spectroscopy. 

\paragraph{Fixed distances.}
We deliberately adopted the geometric distances of \citet{Bailer-Jones2021} as an input, in order to isolate the effect of adding extinction to the target parameters, similar to our analysis in \citet{Ksoll2020}, where extinction was given and only the stellar parameters were inferred. 
These distances are well established and already include the parallax zero-point correction \citep{Lindegren2021}, and for the members of Taurus and Upper Scorpius, at about $140$\,pc, the parallax signal-to-noise ratio is generally high. 
This choice has, however, two consequences. 
First, an error on the distance shifts all six absolute magnitudes by the same amount, which the network, trained with exact distances, interprets as a change in luminosity and compensates in part by a change in extinction through the luminosity--extinction degeneracy. 
This is consistent with the dependence of the map noise on $\SNRPi$ (Fig.~\ref{fig:cINN_extinction_map_vs_snr}) and with the decrease of the luminosity discrepancy with respect to GSP-Phot from about $1$\,dex to about $0.1$\,dex for $\SNRPi>20$ (Fig.~\ref{fig:delta_av_vs_delta_logL_cinn_vs_gspphot}). 
Second, for sources with $\SNRPi\lesssim5$, which dominate the faint field population, the adopted distance is largely determined by the Galactic prior of \citet{Bailer-Jones2021}, which is independent of the photometry and extinction of the star. 
Both methods that we compared against, however, inferred the distance jointly with the stellar parameters, doing the same is part of our ongoing development.

Among the discussed effects that may contribute to the discrepancies between our results and that of the other studies, we have so far not identified a clearly dominating factor, although distance and photometric uncertainties appeared as likely candidates.
It is, however, also possible that the observed offsets might be caused by a compound effect of all these different aspects that are currently unaccounted for in our approach.
As it is evident that the discussed effects should not be neglected in a comprehensive analysis tool and since most of them are part of subsequent development strategy, an even more in-depth analysis of the causes of the extinction discrepancies falls outside of the scope of this study.

\subsection{Stellar ages}
\label{sec:discussion_stellar_ages}
For the final point of this discussion, we want to briefly touch upon the topic of the stellar ages.
As we have outlined, our cINN does not only predict extinction, but jointly estimates the stellar properties, including stellar age and mass. % in a single prediction step as well.
Neither GSP-Phot nor \cite{Zhang2025} predicted ages in their analysis, so we did not conduct any detailed comparisons here.
In general, independently and reliably labelled data is hard to come by in this case.
Nevertheless, there are studies that have published ages for individual young cluster members of both Taurus and Upper Scorpius, such as \cite{Liu2021} or \cite{MiretRoig2022}, respectively.
Cross-matching our predictions with these catalogues of bona fide cluster members, we also found discrepancies in our predicted ages for Taurus and Upper Scorpius.
In particular, it appeared that the cINN is overestimating the ages of both clusters in its current setup by about an order of magnitude. 
For the field population, we have also observed an unusual behaviour, with MAP estimates peaking at $\logage \approx 8.4$ (for stars with $\SNRPi > 20)$.
In the latter case, this appeared to be a result of spurious MAP estimates derived from very wide and flat age posterior distributions, however, rather than a systematic offset.
Wide and uncertain age posteriors are to be expected in these cases, as the field consists predominantly of old, low-mass main-sequence stars, which are notoriously hard to age-date from photometry alone \citep{Soderblom2010}.
For the younger population on the other hand, it is unclear at this stage whether these discrepancies occur because of the simultaneous joint estimation of age and extinction or if they are likewise related to one or more of the currently unaccounted for effects, including the known systematics of pre-main-sequence models \citep{Herczeg2015, Feiden2016}. 
For reference, in \cite{Ksoll2020}, we demonstrated that the recovery of the age of a young cluster is feasible with our approach when extinction is given as an input with HST photometry.
A more detailed analysis of the stellar ages (and potential biases therein) will be part of our follow-up development.

\section{Summary and conclusions}
\label{sec:summary}

In this study, we extended the simulation-based cINN approach of \citet{Ksoll2020} to estimate extinction jointly with the stellar parameters from Gaia and 2MASS photometry alone. Our main conclusions are:
\begin{enumerate}
    \item We built a flexible forward model, based on our own interpolation of the PARSEC v1.2S tracks and on-the-fly photometry synthetisation, which avoids discretisation artefacts in age and extinction and allows the priors, the extinction law and the photometric bands to be changed without regenerating model grids.
    \item On noise-free synthetic data, the cINN recovered $A_V$ with an RMSE of 0.16\,mag, \logteff with 0.07\,dex and \logl with 0.2\,dex. Stellar age remained the hardest parameter (0.56\,dex) to constrain, owing to the intrinsic degeneracy for low-mass main-sequence stars, which the posteriors capture. The posteriors themselves appeared well calibrated with mean absolute calibration errors in the range of $1-4\%$.
    \item Applied to Taurus and Upper Scorpius, the cINN recovered the same extinction structures as GSP-Phot and \citet{Zhang2025}, on the sky and as a function of distance.
    \item The cINN extinctions are systematically higher by 0.5--0.7\,mag. The offset is present for foreground stars and is not removed by an extinction prior concentrated at zero; our diagnostics point to the neglected distance uncertainties, which the network compensates through the luminosity--extinction degeneracy.
    \item $11-13\%$ of the sources received no solution after posterior sample rejection and our extinction maps are comparatively noisy for low \SNRPi, which we trace to the neglected photometric and distance uncertainties.
    \item The approach provides amortised posteriors at $29.8\pm0.1$\,ms per star (for 4096 posterior samples on an NVIDIA RTX 2080Ti), reaches red and strongly reddened sources beyond the published XP spectra, and yields ages and masses jointly with extinction. In its Gaia+2MASS configuration, its depth is set by 2MASS for most stars, and all six bands are required.
    \item Our subsequent development will focus on noise-aware training, joint inference of distances, physically motivated population (age and mass) and extinction priors, and a free metallicity and extinction-law slope.
\end{enumerate}

\begin{acknowledgements}
    The team at Heidelberg University acknowledges financial support from the European Research Council via the ERC Synergy Grant ``ECOGAL'' (project ID 855130),  from the German Excellence Strategy via the Heidelberg Cluster  ``STRUCTURES'' (EXC 2181 - 390900948), from the German Ministry for Economy and Energy (BMWE) in project ``MAINN'' (funding ID 50OO2206), and  from the German and French Science Foundations, DFG and ANR, in project ``STARCLUSTERS'' (funding ID KL 1358/22-1). . 
    They are also grateful for computing resources provided by the Ministry of Science, Research and the Arts (MWK) of the State of Baden-W\"{u}rttemberg through bwHPC and the German Science Foundation (DFG) through grants INST 35/1134-1 FUGG and 35/1597-1 FUGG, and also for data storage at SDS@hd funded through grants INST 35/1314-1 FUGG and INST 35/1503-1 FUGG. 
    V.K. and P.G. also acknowledge financial support by the Carl-Zeiss-Stiftung.
    Part of this work was supported by the German \emph{Deut\-sche For\-schungs\-ge\-mein\-schaft, DFG\/}.
    This work has made use of data from the European Space Agency (ESA) mission {\it Gaia} (\url{https://www.cosmos.esa.int/gaia}), processed by the {\it Gaia} Data Processing and Analysis Consortium (DPAC, \url{https://www.cosmos.esa.int/web/gaia/dpac/consortium}). Funding for the DPAC has been provided by national institutions, in particular the institutions participating in the {\it Gaia} Multilateral Agreement.
\end{acknowledgements}

\bibliographystyle{aa}
\bibliography{./bibtex}

\appendix
\nolinenumbers

\section{Comparison methods}
\label{app:other-tools}
\subsection{GSP-Phot}
\label{app:gpspphot}
The General Stellar Parameterizer from Photometry (GSP-Phot) is part of the astrophysical parameters (APs) inference system \citep[Apsis;][]{Bailer-JonesEtAl2013_Apsis, CreeveyEtAl2023_Apsis} and is designed to infer astrophysical stellar parameters, distances, and extinctions using a Markov Chain Monte Carlo (MCMC) Bayesian forward-modelling approach to simultaneously fit Gaia low-resolution spectra XP (RP/BP), parallax, and apparent G magnitude \citep{LiuEtAl2012_GSP-Phot, AndraeEtAl2023_GSP_Phot}, see also \href{https://gea.esac.esa.int/archive/documentation/GDR3/Data_analysis/chap_cu8par/sec_cu8par_apsis/ssec_cu8par_apsis_gspphot.html}{GSP-Phot documentation}. 
At the heart of GSP-Phot is \texttt {Aeneas} (q-method), a Bayesian MCMC engine 
that samples the posterior probability density function of the APs given the measured Gaia data (XP spectra, apparent G magnitude and parallax).
It employs an Affine-Invariant Ensemble Sampler \citep{GoodmanEtAl2010_MCMC_AffineInvariance}, which is an ensemble MCMC method, to provide an efficient sampling unaffected by linear transformations of parameters and robust to correlated posterior shapes. \texttt {Aeneas} uses the \texttt{emcee} Python implementation \citep{Foreman-MackeyEtAl2013_emcee} with 4 optimization (fit) parameters, which are \logage, \logmini, metallicity [M/H], and extinction $A_0$, where the latter is a monochromatic extinction parameter at $541.4$nm as used in the extinction law of \citet{Fitzpatrick1999}, assuming constant $R_0 = 3.1$. 
All other astrophysical parameters are derived from these fit parameters and stellar evolutionary, atmospheric and extinction models. 

\texttt{Aeneas} first takes the PARSEC 1.2S Colibri isochrone grids and uses 3D linear interpolation over \logage, \logmini, and metallicity [M/H] to obtain self-consistent temperature \teff, surface gravity \logg, radius $R$ and absolute magnitude $M_G$. 
From the temperature, gravity, metallicity and extinction $A_0$, {\texttt Aeneas} then uses 4D linear interpolation to obtain separately four synthetic XP spectra using a set of model grids of synthetic spectral energy distributions (SEDs) that cover the whole range of stellar spectral types \citep{CreeveyEtAl2023_Apsis} based on 4 different stellar atmosphere models:  MARCS \citep{GustafssonEtAl2008_MARCS}; PHOENIX \citep{BrottEtAl2005_PHOENIX, Husser2013_PHOENIX}; A \citep{Kochukhov2005_LLModels_A}; OB \citep{BouretEtAl_TLUSTY_OB}.

To reduce the computational cost of the MCMC method, \texttt{Aenas} starts with initial parameters that are within a ball around a preliminary set of optimisation parameters as obtained from machine learning algorithms trained on synthetic SEDs: \texttt {Extremely Randomised Trees} \citep[\texttt ExtraTrees;][similar to a random forest; implemented in \texttt {scikit-learn}]{GeurtsEtAl2006_ExtraTrees} or/and the Bayesian forward gradient-descent algorithm \texttt {ILIUM} \citep{Bailer-Jones2010_ILIUM}.

As usual in a Bayesian approach, the posterior probability in \texttt {Aeneas} is sampled using priors and a log-likelihood computed as the sum of three $\chi^2$ statistics, hence assuming Gaussian uncertainties. 
The priors have been preferentially defined here not only as traditionally for the actual fit parameters (extinction) but also partly for the derived parameters based on the HR diagram.
The prior for extinction $A_0$ is restricted to the range $A_0 \in [0, 10]$ and follows an ad-hoc exponential form, $P(A_0|D \propto \exp(-A_0/\mu)$, where the mean value $\mu$ depends on Galactic latitude and distance $D$.
The priors for $A_G$ and the radius $R$ are delta distributions that set the values respectively to the extinction obtained from integrating the SED and to the value provided by the isochrone.
The distance prior follows a decreasing exponential $P(D) \propto  D^2 \exp({-D/L})$ \citep{Bailer-Jones2015_DistancePrior_Parallax}, where the length scale $L$ depends on Galactic coordinates and has been mapped from the Gaia EDR3 mock catalogue of \citet{RybizkiEtAl2020_GaiaDR3_MockCatalog}. 
We note that the choice of this mock catalogue leads to a lack of sources at large distances, resulting in a systematic underestimation of distances by GSP-Phot. 
And finally, they use a Hertzsprung-Russell-diagram prior to define the probability distribution $P([M/H],T_{\rm eff},\log g,M_G)$ constructed from the Gaia Universe Model Snapshot \citep[GUMS;][]{Robin2012}.

As said above, the likelihood is defined with three $\chi^2$ statistics. 
The first, $\chi_{XP}^2$, is associated to the quadratic difference of observed and synthetic XP spectra.
The second, $\chi_{G}^2$, is linked to the difference between the apparent G magnitude observed and the synthetic apparent G magnitude obtained using $A_G$ extinction and an estimate of the distance $D$ computed from the apparent flux $F$ of XP spectra ($F=\left ({R}/{D}\right )^2 \sigma_B T_{\rm eff}^4$).
The last, $\chi_{\pi}^2$, is computed from the observed astrometric parallax $\pi$ and the parallax estimated from the distance D $\pi_{e}=1/D$.

The posterior probabilities of the stellar astrophysical parameters $T_{\rm{eff}},\,\log g,\, [M/H],\,A_0,A_G,M_G$ and distance of the star $D$ are sampled for each of the four atmospheric models; the best library is identified as the one having the highest mean log-posterior value averaged over the MCMC samples. 
As a final result, the astrophysical stellar parameters $T_{\rm eff},\log g,[M/H],A_0,A_G,R,M_G$ and the distance $D$ are given as their expected value (mean values).

\section{Additional material - Synthetic tests}
\label{app:synth}

This appendix provides extra material and additional analysis for the application of the cINN on synthetic test data.

\subsection{Outlier locations in the Hertzsprung-Russell diagram}
\label{app:synth_outliers}

\begin{figure*}
    \centering
    \includegraphics[width=\linewidth]{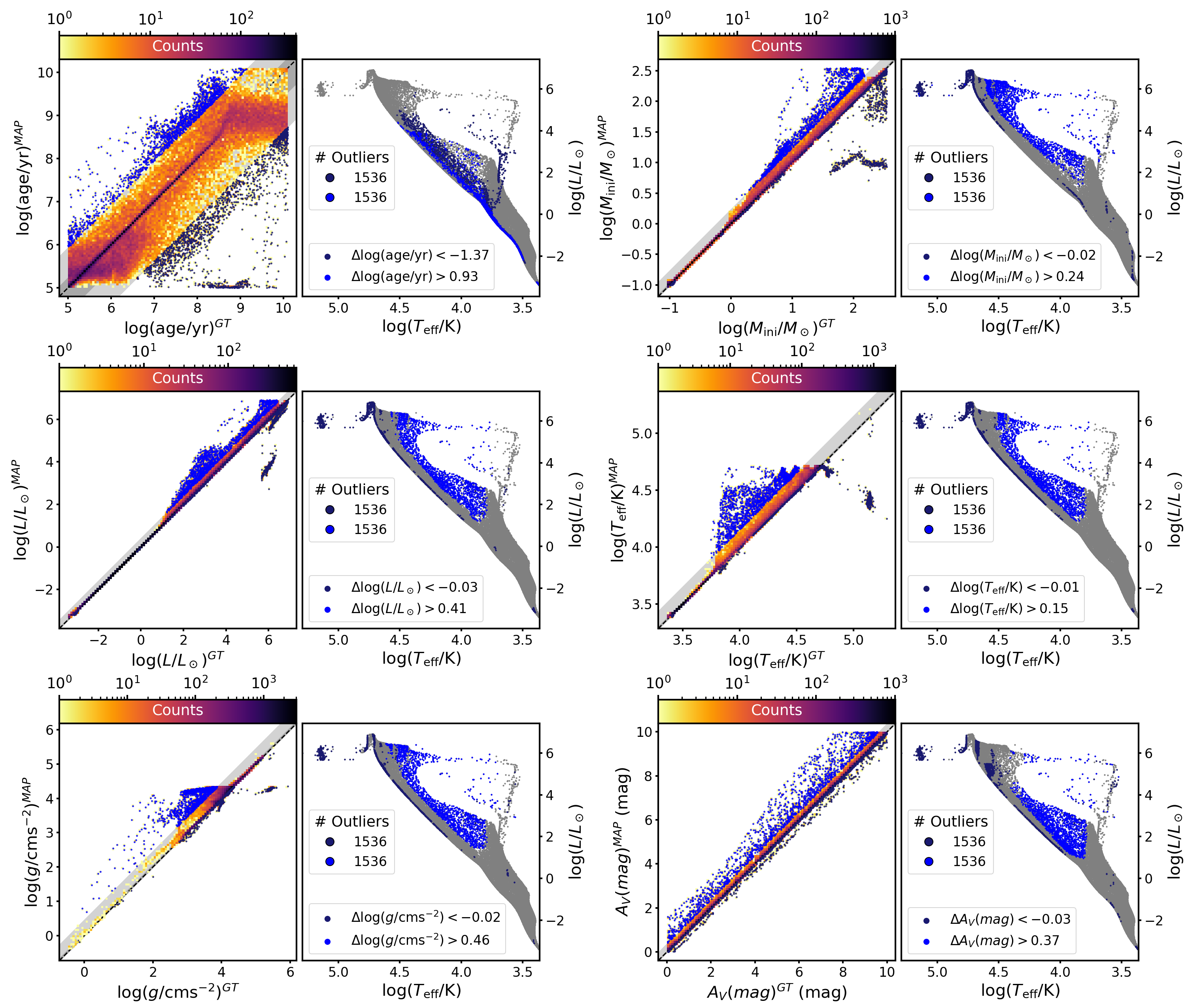}
    \caption{Outlier analysis of the MAP estimates on the synthetic test data. In each pair of panels, the left diagram shows the 2D histogram comparing the MAP estimates to the ground truth (as in Fig.~\ref{fig:MAP_vs_GT_1to1}) with outliers highlighted, while the right diagram indicates the position of these outliers in the HRD. Cases of notable underestimation (below the 2.5\% percentile of the residuals) are denoted in purple and cases of overestimation (above the 97.5\% percentile of the residual) in blue.}
    \label{fig:synth_hrd_outliers}
\end{figure*}

Expanding upon Fig.~\ref{fig:MAP_vs_GT_1to1}, Fig.~\ref{fig:synth_hrd_outliers} highlights where outliers in the 1-to-1 comparison of cINN MAP estimates and ground truth are located in the Hertzsprung-Russel diagram. 
In particular, we analysed the predictions that fall outside of the 95\% percentile interval (i.e.~the interpercentile range from the $2.5$\% to $97.5$\% percentile) of the residual $X^\mathrm{MAP} - X^\mathrm{GT}$, distinguishing between cases of underestimation and overestimation.
We note that due to the overall skewness of the residual distributions the magnitude of the discrepancy in the under- and overestimation cases may be significantly different. 
In particular, the outliers that underestimate tend to have a tiny absolute error with respect to the ground truth for all parameters except for the age.
Comparing to the prior distributions in the HRD shown in Fig.~\ref{fig:Sampling_TS_1}, this figure demonstrates that the largest outliers lie predominantly in regions of the HRD that are less well sampled in our training data set.
The primary exception is the age prediction, where we also identified a large set of outliers on the main-sequence, which is a result of the fact that determining age for these types of sources from photometry alone is highly difficult. 

\subsection{Full posterior distributions}
\label{app:synth_posteriors}
\begin{figure*}
    \centering
    \includegraphics[width=\linewidth]{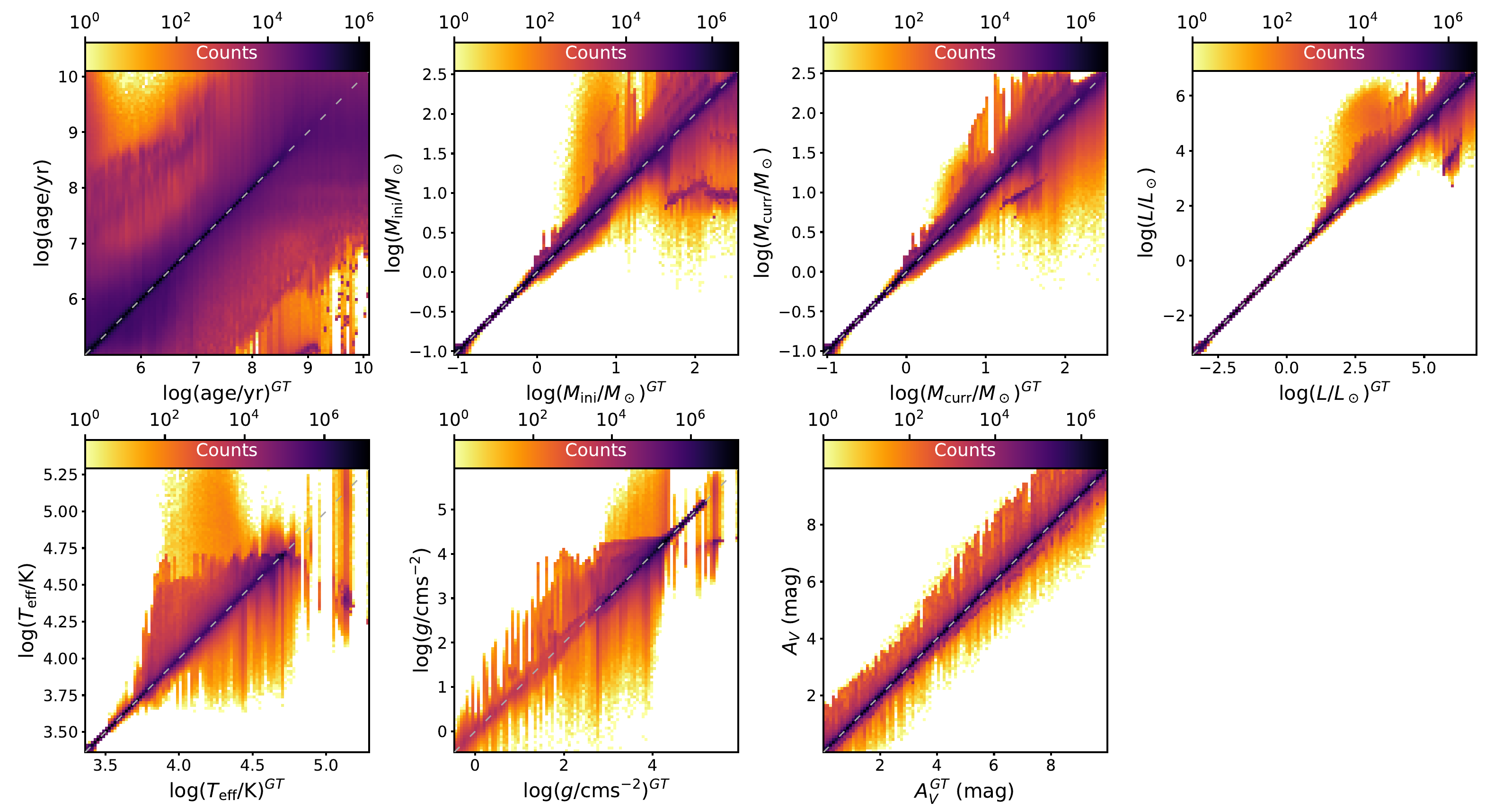}
    \caption{2D histogram comparison of the cINN posterior distributions to the ground truth for $61\,440$ synthetic test examples. The dotted grey line indicates a perfect one-to-one correlation. We note that the colour bar is logarithmically scaled to increase the contrast.}
    \label{fig:posterior_vs_gt}
\end{figure*}

Complementary to Fig.~\ref{fig:MAP_vs_GT_1to1}, we present in Fig.~\ref{fig:posterior_vs_gt} a comparison of the entire predicted posteriors to the ground truth for the $61\,440$ synthetic test samples.
This figure further demonstrates the difficulty in predicting stellar age, highlighting the large average widths of the age posteriors due to degeneracy, whereas the posterior components of the other target parameters appear on average much more constrained.

\subsection{Resimulation}
\label{app:synth_resimulation}
\begin{figure*}
    \centering
    \includegraphics[width=0.8\linewidth]{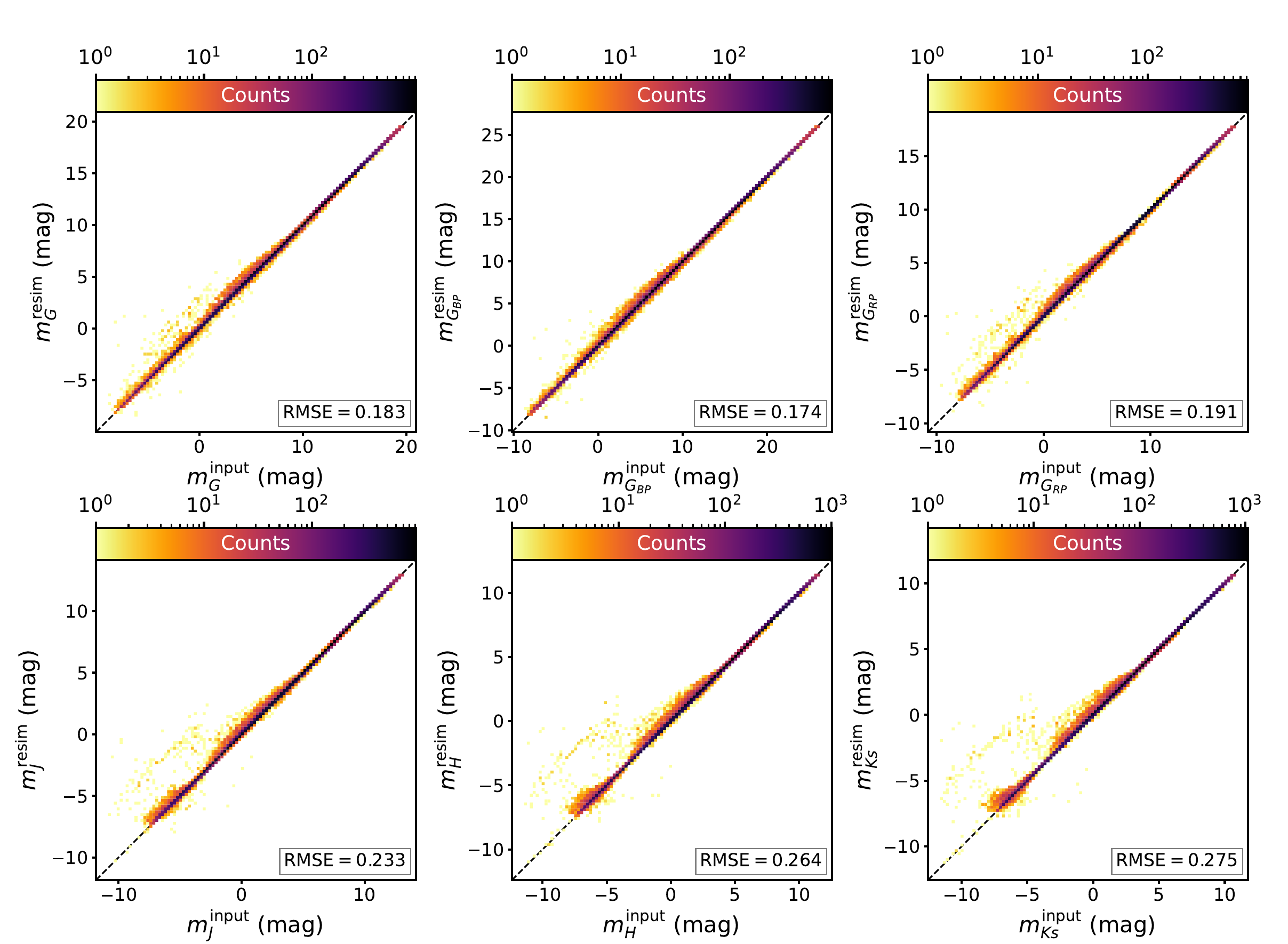}
    \caption{2D histogram comparison of the resimulated photometry to the input photometry for $61\,440$ synthetic test examples. The resimulated photometry was interpolated using our full synthetic model interpolation pipeline based on the MAP estimates of age, initial mass and extinction from our network. The black dashed line indicates a perfect 1-to-1 match. We emphasise that the colour-bar is logarithmically scaled for enhanced contrast.}
    \label{fig:resim_testset}
\end{figure*}

To further verify the self-consistency of the MAP predictions of our cINN on the synthetic test data (see Fig.~\ref{fig:MAP_vs_GT_1to1}), we performed a resimulation experiment.
This means that we took the MAP estimates of \logage, \logmini and $A_V$ and reinterpolated the corresponding stellar evolutionary models and synthetic photometry using our full interpolation pipeline.
We compared this resimulated photometry for the test data to the input synthetic photometry to validate that the MAP estimates correspond to models with compatible photometry.
Figure~\ref{fig:resim_testset} summarises this experiment and reveals and overall good agreement, with RMSEs between $0.28$\,mag in the Ks-band to $0.17$\,mag in the $\mathrm{G_{BP}}$ band.
This, in particular, identified the underestimated ages of the old synthetic test examples (i.e.~$\logage > 8.5$) as degenerate solutions rather than mispredictions.

\subsection{Additional material - Application to Taurus and Upper Scorpius}
\label{app:taurus_usco}

This appendix provides extra material and additional analysis for the application of the cINN to the observational data of the Taurus and Upper Scorpius star-forming regions.

\subsubsection{Photometric errors of rejected sources}
\label{app:taurus_usco_resim}

\begin{figure*}
    \centering
    \includegraphics[width=\linewidth]{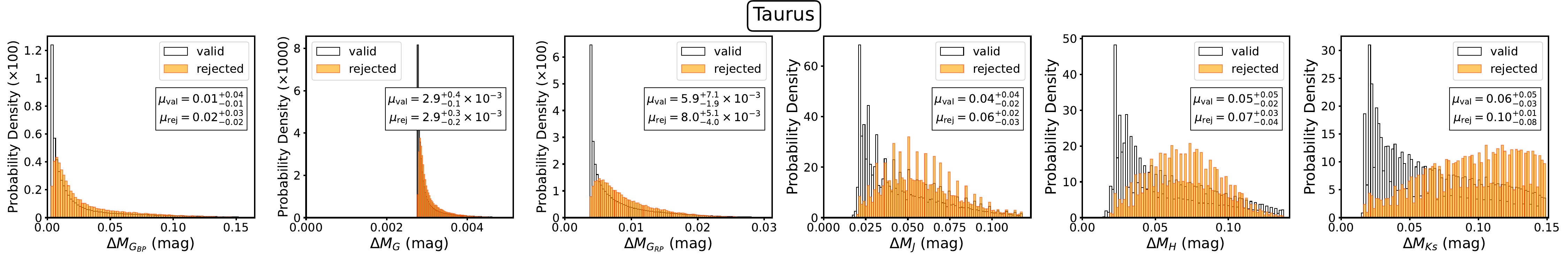}
    \includegraphics[width=\linewidth]{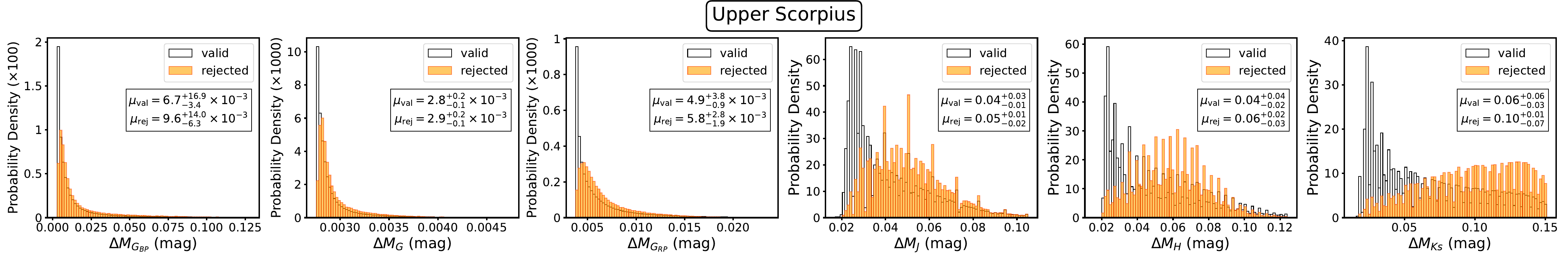}
    \caption{Histograms of the Gaia and 2MASS photometric uncertainties for the Taurus and Upper Scorpius datasets. The white histograms represent the sources, for which our cINN approach returns valid posterior estimates, whereas the orange histograms indicate sources, where all generated posterior samples were rejected in our posterior analysis. For convenience, each panel also lists the median (and 16\%/84\% percentile) of the photometric uncertainty for both the subsets with valid cINN predictions ($\mu_\mathrm{val}$) and completely rejected posterior distributions ($\mu_\mathrm{rej}$). }
    \label{fig:sample_rejection_vs_phot_errs}
\end{figure*}

In our application of the trained cINN to the photometric data Taurus and Upper Scorpius in Section~\ref{sec:results_taurus_usco}, we identified a notable fraction of query sources (about $13\%$ in the Taurus and $11\%$ in the Upper Scorpius dataset), where our evaluation pipeline did not return results for the stellar properties and extinction.
These cases did not constitute catastrophic failures of our model, but resulted from our strict posterior rejection scheme, which did not allow the cINN to extrapolate at all as described in Section~\ref{sec:MAP_sample_rejection}.
To identify potential reasons, why the cINN needed to extrapolate for these query sources, we investigated the uncertainties of the input photometry, comparing between sources with valid cINN predictions and without.
We note that the Gaia DR3 catalogue does not directly list photometric uncertainties, but rather the flux uncertainties, so that we computed the former via error propagation
\begin{equation}
    \Delta M_X = \sqrt{\left(\frac{-2.5}{\ln 10} \frac{\Delta F_X}{F_X}\right)^2 + \left(\Delta \mathrm{ZP}_{X}\right)^2},
\end{equation}
where $F_X$ and $\Delta F_X$ denote the flux and flux error in band $X$, respectively, while $\Delta\mathrm{ZP}_{X}$ represents the uncertainties of the Gaia photometric zero points \citep[taken from][]{Riello2021}. 
Figure~\ref{fig:sample_rejection_vs_phot_errs} summarises this investigation and indicates that the on average larger photometric uncertainties in the 2MASS bands could be a potential cause for the extrapolation of the cINN and subsequent total posterior rejection for these query sources.
As our cINN is currently not accounting for photometric uncertainties, particularly uncertain input photometry may fall outside the learned input feature space, triggering the undesired extrapolation.
Notably, we did not find a similar correlation of the sources without cINN prediction and the distance uncertainty \SNRPi.

\subsubsection{Extinction maps for different parallax quality}
\label{app:taurus_usco_extinction_maps_vs_snr}

\begin{figure*}
    \centering
    \includegraphics[width=\linewidth]{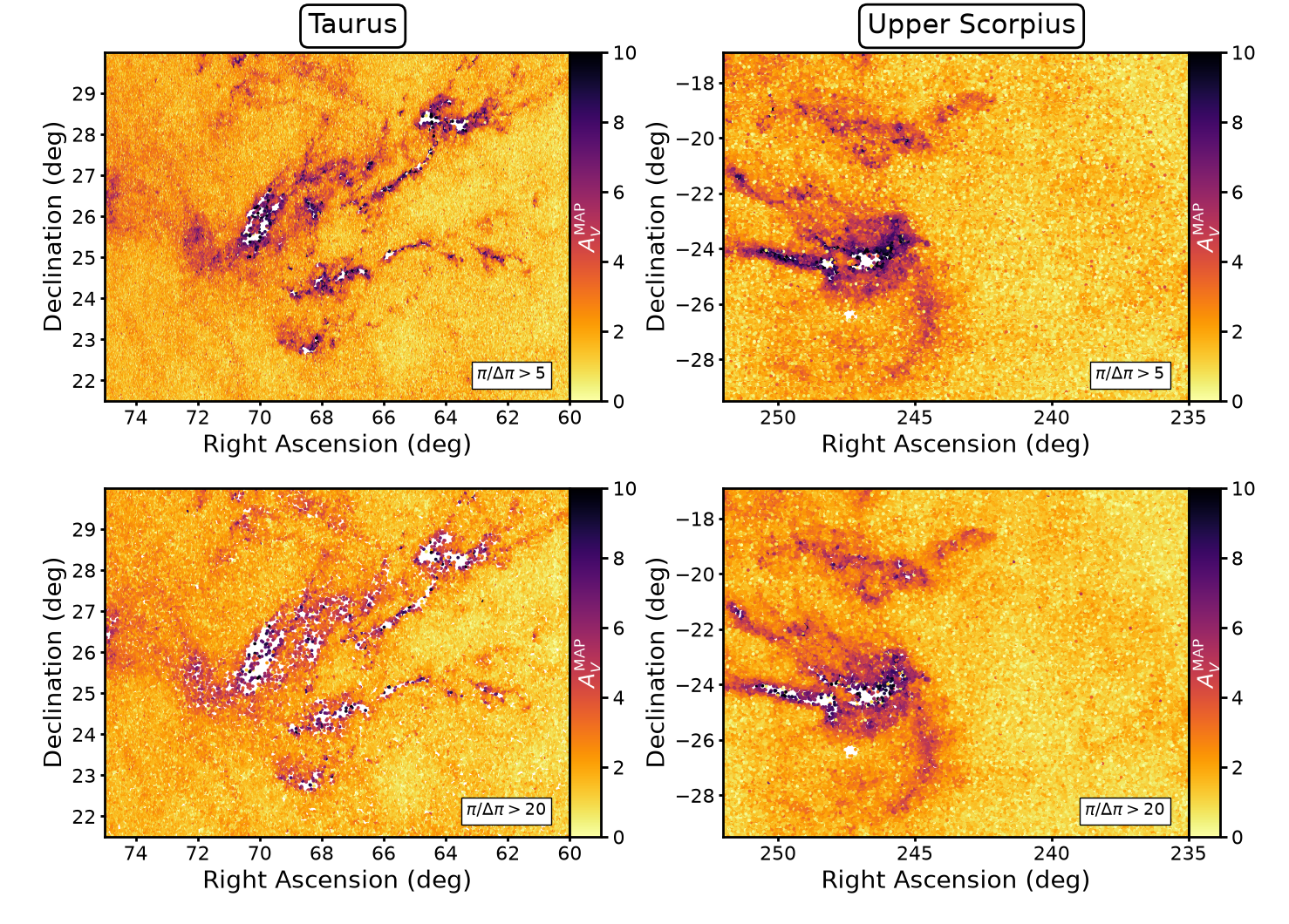}
    \caption{Extinction maps for the star-forming regions Taurus and Upper Scorpius based on MAP extinction estimates from the cINN. Complimentary to Fig.~\ref{fig:cinn_exmap}, each row shows the cINN extinction map for a different threshold on the S/N (\SNRPi) for the Gaia parallax. Each panel shows a scatter plot, colour-coded according to $A_V^\mathrm{MAP}$. }
    \label{fig:cINN_extinction_map_vs_snr}
\end{figure*}

Figure~\ref{fig:cinn_exmap} in the main paper presents the extinction maps derived from our cINN predictions for the star-forming regions Taurus and Upper Scorpius.
In these maps, we identified a notable amount of noise, which appears to correlate with the quality of the query data, in particular the distance uncertainty \SNRPi.
Figure~\ref{fig:cINN_extinction_map_vs_snr} highlights this correlation by demonstrating that the noise notably decreases for higher thresholds on \SNRPi.
The noise itself is likely a result of the unaccounted-for distance uncertainties in our current cINN model.

\subsubsection{Resimulation for observational data}
\label{app:taurus_usco_resimulation}

\begin{figure*}
    \centering
    \includegraphics[width=1\linewidth]{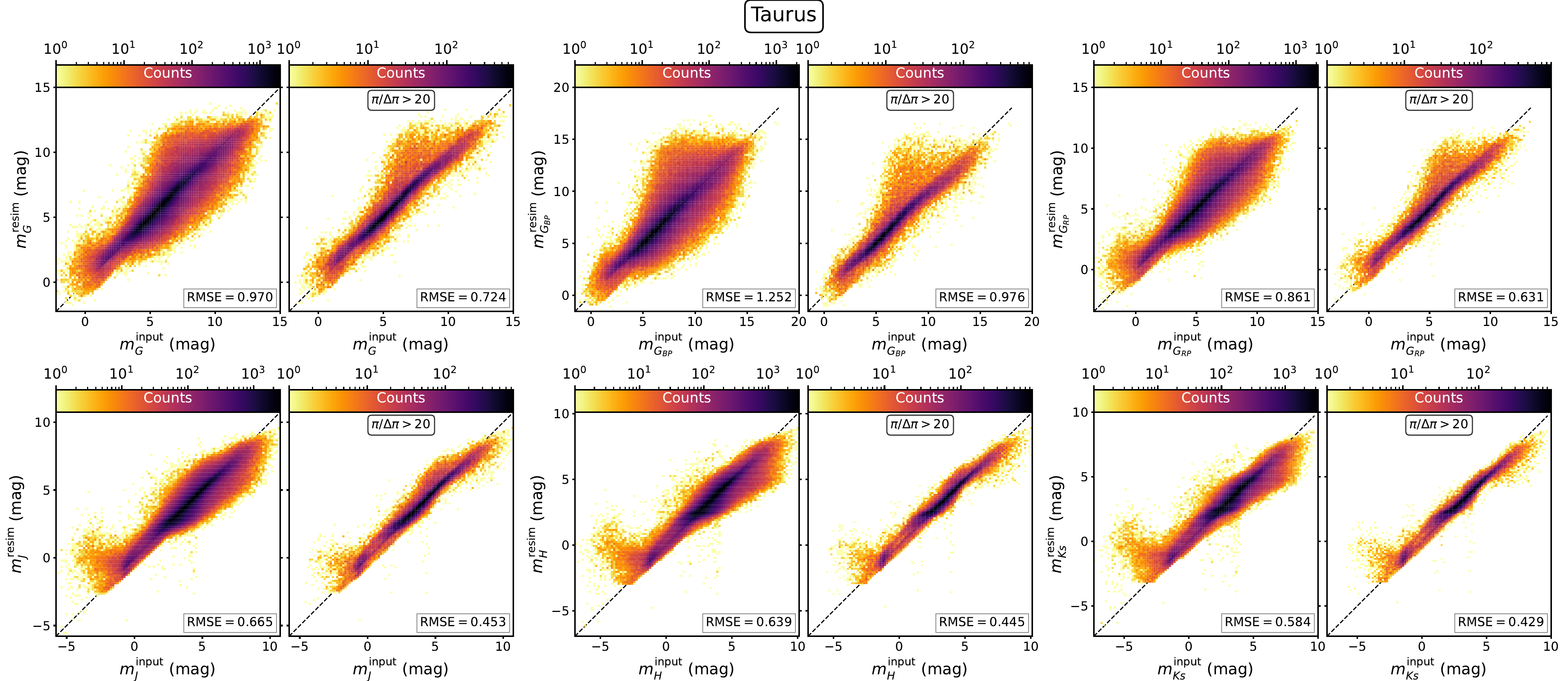}
    \includegraphics[width=1.0\linewidth]{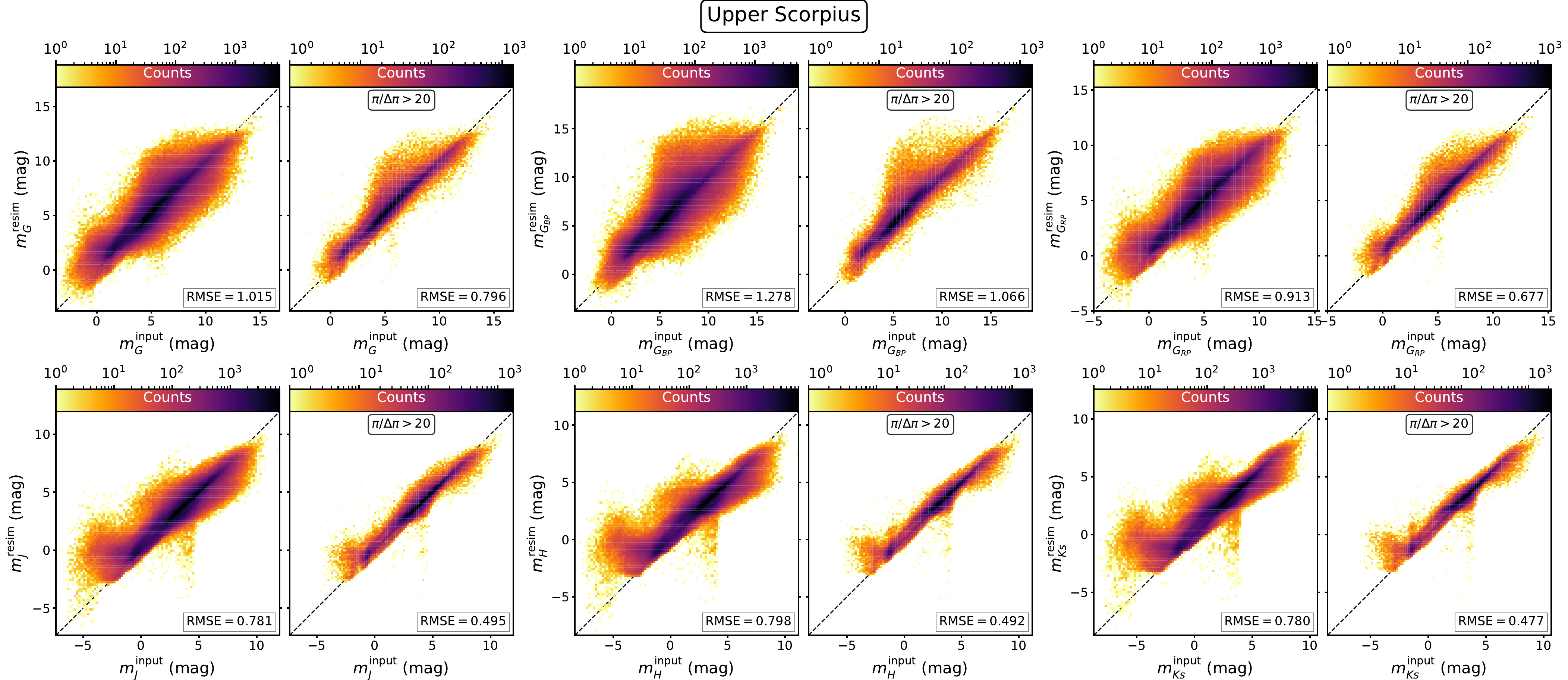}
    \caption{2D histogram comparison of the resimulated photometry to the input photometry for the observational datasets of Taurus and Upper Scorpius. Shown are the resimulation results based on the MAP estimates for age, initial mass and extinction. The black dashed line indicates a perfect 1-to-1 recovery. We note that the colour-bar is scaled logarithmically for increased contrast. In each group of panels, the left panel indicates the result for the entire dataset, whereas in the right panel a S/N threshold of $\pi/\Delta\pi > 20$ is applied on the Gaia parallax to highlight the outcome on the subset with the best quality.}
    \label{fig:resim_taurus_usco}
\end{figure*}

To further validate our predictions on the observational data of Taurus and Upper Scorpius, we performed an analogous resimulation experiment to our analysis of the synthetic test models.
Figure~\ref{fig:resim_taurus_usco} summarises this experiment and highlights, in particular for the high-quality data (i.e.~$\SNRPi > 20$), an overall agreement of our predictions with the observed input photometry.

\subsubsection{Residuals with respect to the comparison catalogues}
\label{app:taurus_usco_residuals}

\begin{figure*}
    \centering
    \includegraphics[width=\linewidth]{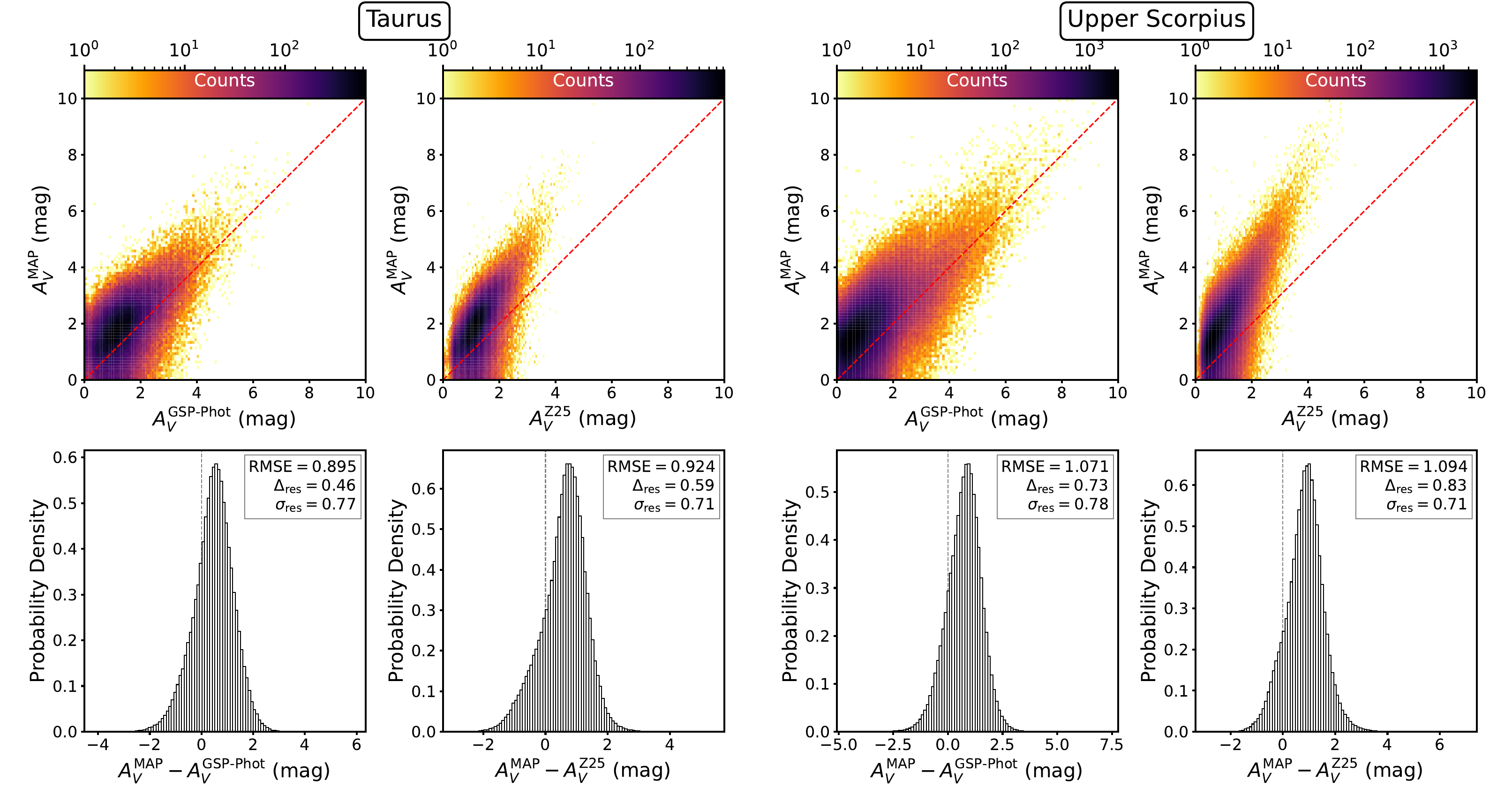}
    \caption{Comparison of the cINN MAP estimates for extinction to the results from GSP-Phot and \cite{Zhang2025} for the Gaia data of Taurus and Upper Scorpius. The top panels show a 2D histogram of a direct 1-to-1 comparison of the cINN estimates to the other two methods. The black dashed line denotes a perfect 1-to-1 recovery. The bottom panels provide a 1D histogram of the corresponding residuals. The grey dashed vertical line marks a zero residual for reference.}
    \label{fig:av_map_vs_gspphot_z25}
\end{figure*}

In the comparison of our extinction maps to the independent spectroscopy-based approaches from GSP-Phot and \cite{Zhang2025}, we identified qualitatively a good recovery of the overall structure of extinction in both Taurus and Upper Scorpius.
Quantitatively, on the other hand, the comparison revealed some systematic offsets, highlighted in Fig.~\ref{fig:av_map_vs_gspphot_z25}.
In particular, with respect to GSP-Phot, we can observe a somewhat fixed offset to higher extinction in our results, whereas compared to \cite{Zhang2025}, there is also a slope in addition to the offset. 
The latter might be a result of \cite{Zhang2025} treating $R_V$ as a free parameter in their analysis, whereas we fixed $R_V=3.1$ for training the cINN, although preliminary experiments with synthetic photometry generated with differing $R_V$ did not reproduce the observed behaviour.

\subsubsection{Luminosity comparison with GSP-Phot}
\label{app:taurus_usco_logl_vs_deltaAv}

\begin{figure*}
    \centering
    \includegraphics[width=\linewidth]{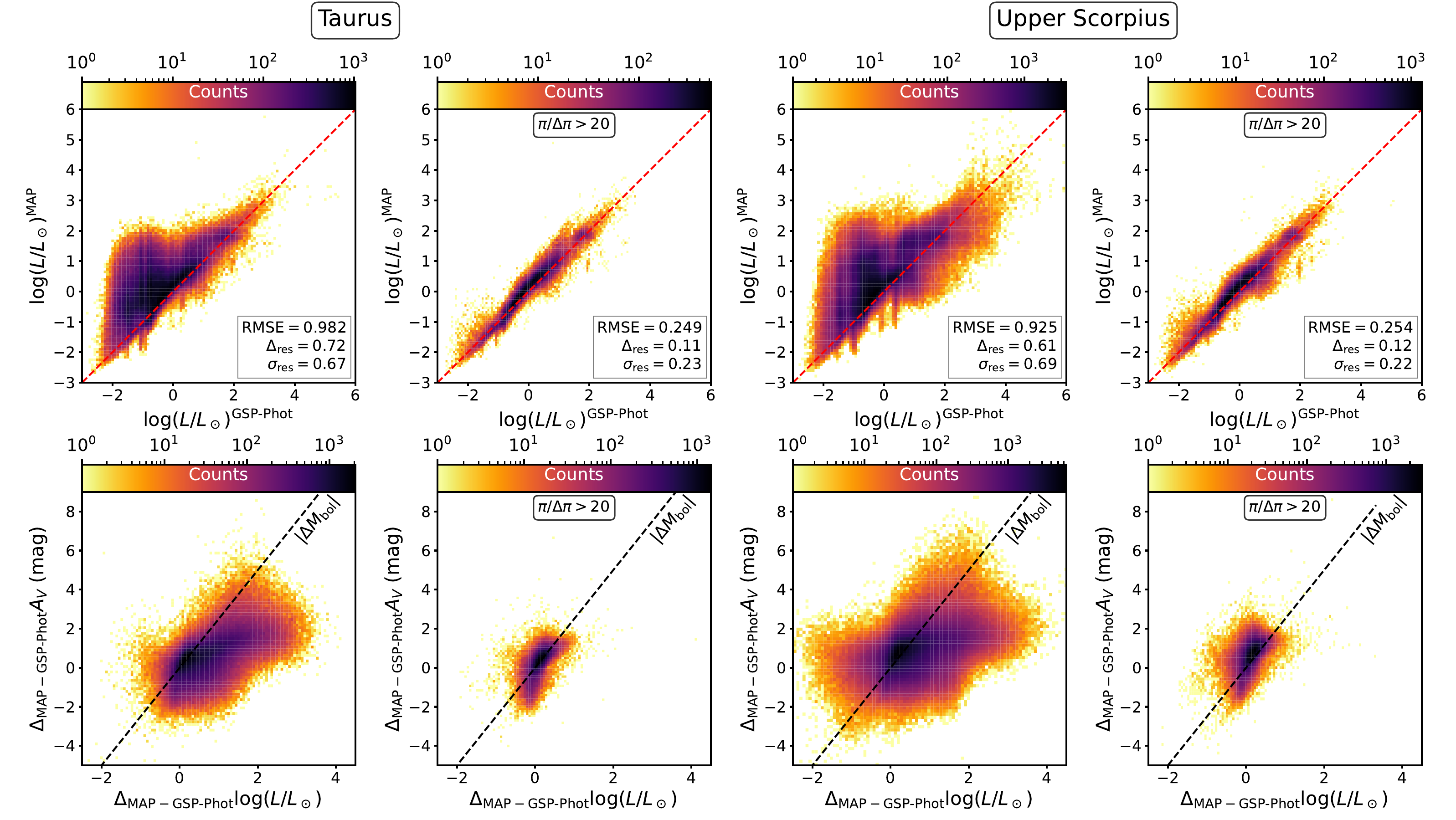}
    \caption{Analysis of extinction discrepancies between the cINN and GSP-Phot for the Taurus and Upper Scorpius Gaia data. The top panels show a 2D histogram comparing the MAP luminosity estimates of the cINN to the GSP-Phot (blackbody) results (as determined from the GSP-Phot estimates for stellar radius and effective temperature). The black dashed line marks a perfect 1-to-1 recovery for reference. The bottom panels present a comparison of the luminosity residuals to the extinction residuals between cINN and GSP-Phot estimate. Here the black dashed line indicates the absolute difference in bolometric magnitude as a function of the luminosity residual, i.e.~$|\Delta M_\mathrm{bol}| = |2.5\Delta\log(L/L\odot)|$ for reference. For both the Taurus and Upper Scorpius panels, the left set of panels shows the result for the entire dataset, whereas the right panels present the subset with a S/N threshold of $\pi/\Delta\pi > 20$ on the Gaia parallax to indicate the outcome for the highest-quality data. }
    \label{fig:delta_av_vs_delta_logL_cinn_vs_gspphot}
\end{figure*}

To further investigate the offsets in our extinction predictions with respect to the GSP-Phot results, we compared the predicted luminosities.
The Gaia DR3 catalogue does not provide luminosities from GSP-Phot directly, but estimates of effective temperature $T_\mathrm{eff}^\mathrm{GSP\text{-}Phot}$ and stellar radius $R_*^\mathrm{GSP\text{-}Phot}$ instead.
Based on these, we computed luminosities for GSP-Phot following
\begin{equation}
    L^\mathrm{GSP\text{-}Phot} = 4\pi (R_*^\mathrm{GSP\text{-}Phot})^2 \sigma_B(T_\mathrm{eff}^\mathrm{GSP\text{-}Phot})^4,
\end{equation}
where $\sigma_b$ denotes the Stefan Boltzmann constant. 
Figure~\ref{fig:delta_av_vs_delta_logL_cinn_vs_gspphot} provides a direct comparison of the predicted luminosities between the cINN and GSP-Phot, as well as a comparison of the residuals in extinction to the residuals in luminosity.
This experiment indicated that the cINN appears to slightly overestimate the stellar luminosity compared to GSP-Phot, and that the corresponding difference in bolometric magnitude appears to be somewhat consistent with the observed discrepancies in extinction (in particular for the high-quality data with $\SNRPi > 20$).
This suggests that the cINN may be compensating overestimated stellar luminosity with increased extinction to fit the input observations. 

\subsubsection{Sources characterised only by the cINN}
\label{app:taurus_usco_cinn_unique}

\begin{figure*}
    \centering
    \includegraphics[width=\linewidth]{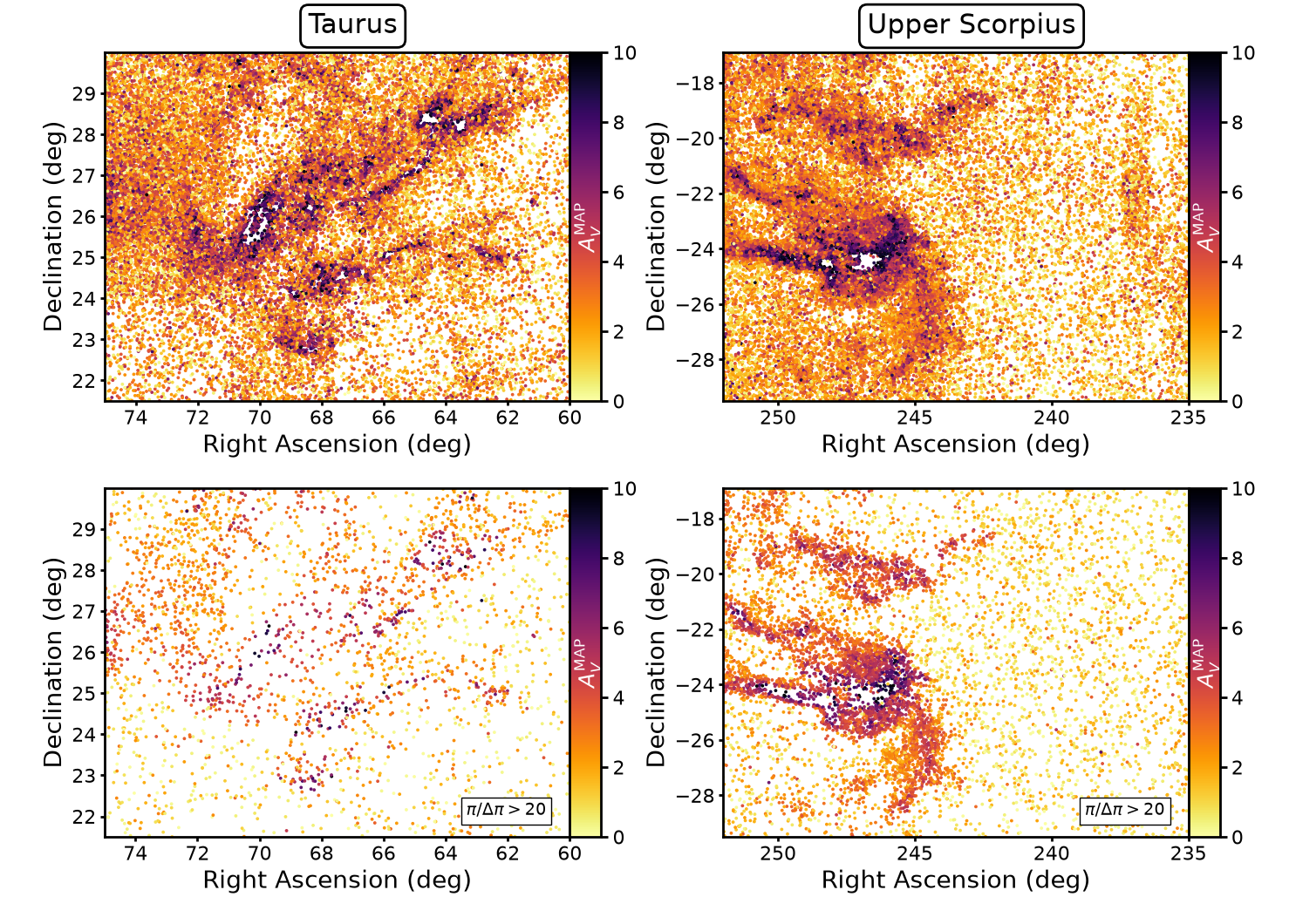}
    \caption{Extinction maps for the star-forming regions Taurus and Upper Scorpius based on the MAP extinction estimates of the cINN, limited to sources for which neither GSP-Phot nor \cite{Zhang2025} provide a result. Analogous to Figs.~\ref{fig:cinn_exmap}, \ref{fig:cinn_exmap_vs_gspphot_z25} and \ref{fig:cINN_extinction_map_vs_snr}, all panels show a scatter plot, colour-coded according to $A_V^\mathrm{MAP}$. Each row presents a different threshold on the Gaia distance uncertainty, i.e.~$\SNRPi$, from now threshold in the top to $\SNRPi > 20$ in the bottom.}
    \label{fig:extinction_map_cinn_unique_results}
\end{figure*}

In our application of the cINN to the observational data, we identified a subset of the data, namely $33\,993$ sources for Taurus and $42\,435$ in Upper Scorpius, where neither GSP-Phot nor \cite{Zhang2025} provided an estimate of extinction.
Figure~\ref{fig:extinction_map_cinn_unique_results} provides a visualisation of these sources with a unique extinction estimate only from the cINN.
Most of this data is relatively low quality (more than 50\% has $\SNRPi < 5$), so that the cINN extinction estimates may need to be treated with caution, and results in a noisy extinction map.
Nevertheless, focusing on the high quality data ($\SNRPi > 20$) revealed a notable concentration of these sources in the denser (optical) cloud environments of these star-forming regions with our extinction estimates demonstrating rather consistent extinction gradients.
This may indicate that our photometry-only approach was able to probe these high-extinction areas to a deeper degree than the other two spectra-reliant studies.
However, from the Gaia flags alone, we could not identify why GSP-Phot did not provide solutions for these sources. 

\end{document}